\documentclass[final,1p,times,authoryear]{elsarticle}

\usepackage{amssymb}
\usepackage{amsthm}

\usepackage{graphicx}
\usepackage{subfig}

\usepackage{url}
\usepackage{lineno}

\newcommand{\approxprop}{\mathrel{\vcenter{
      \offinterlineskip\halign{\hfil$##$\cr
          \propto\cr\noalign{\kern2pt}\sim\cr\noalign{\kern-2pt}}}}}

\journal{International Journal of Human-Computer Studies}

\begin{document}

\begin{frontmatter}



\title{A Monte Carlo Simulation Approach for Quantitatively Evaluating  Keyboard Layouts for Gesture Input}

\author[label1]{Rylan T. Conway}
\author[label2]{Evan W. Sangaline}
\address[label1]{Physics Department, University of California, Davis, CA}
\address[label2]{National Superconducting Cyclotron Laboratory, Michigan State University, East Lansing, MI}

%\author{}

%\address{}

\begin{abstract}
Gesture typing is a method of text entry that is ergonomically well-suited to the form factor of touchscreen devices and allows for much faster input than tapping each letter individually.
The QWERTY keyboard was, however, not designed with gesture input in mind and its particular layout results in a high frequency of gesture recognition errors.
In this paper, we describe a new approach to quantifying the frequency of gesture input recognition errors through the use of modeling and simulating realistically imperfect user input.
We introduce new methodologies for modeling randomized gesture inputs, efficiently reconstructing words from gestures on arbitrary keyboard layouts, and using these in conjunction with a frequency weighted lexicon to perform Monte Carlo evaluations of keyboard error rates or any other arbitrary metric.
An open source framework, Dodona, is also provided that allows for these techniques to be easily employed and customized in the evaluation of a wide spectrum of possible keyboards and input methods.
Finally, we perform an optimization procedure over permutations of the QWERTY keyboard to demonstrate the effectiveness of this approach and describe ways that future analyses can build upon these results.
\end{abstract}

\begin{keyword}
touchscreen keyboards \sep gesture input \sep model-based design \sep Monte Carlo simulation



\end{keyword}

\end{frontmatter}

%\linenumbers

\section{Introduction}
\begin{figure}[tbh]
\begin{centering}
\includegraphics[scale=0.65]{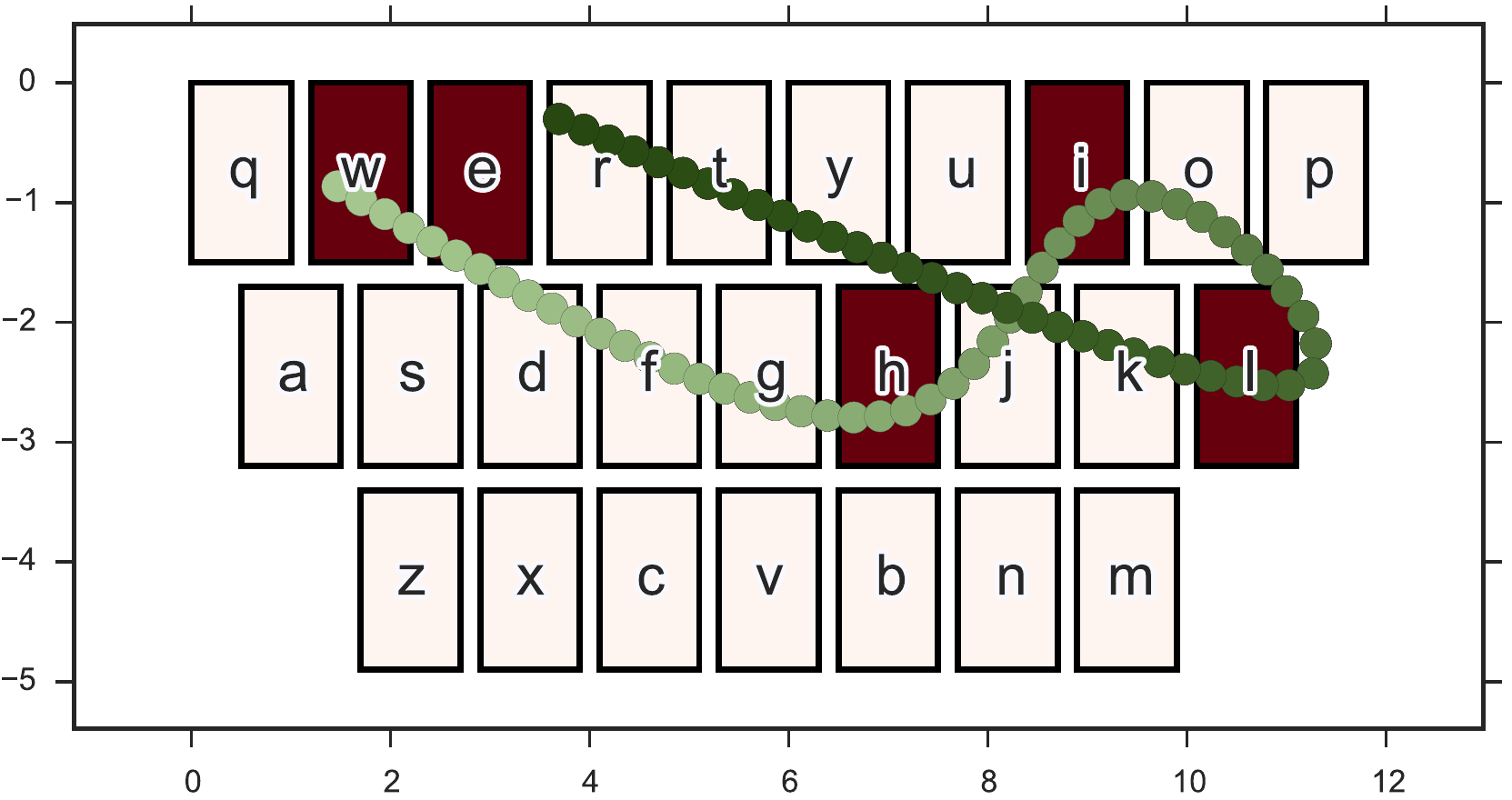}
\par\end{centering}
\protect\caption{A gesture input collision between the words ``while'' and ``whole''. The
gesture input pattern, represented by the series of green markers, was intended
to represent the word ``whole'' but instead was incorrectly matched
with the word ``while''.  \label{fig:Example-swipe-collision}}
\end{figure}

The advent of smartphones and tablets has made the use of touchscreen
keyboards pervasive in modern society. However, the ubiquitous QWERTY
keyboard was not designed with the needs of a touchscreen keyboard
in mind, namely accuracy and speed. The introduction of gesture or stroke-based 
input methods significantly increased the speed that text could be
entered on touchscreens [\cite{Montgomery,SHARK,shapeWriter,swype}]. 
However, this method introduces some new problems that can occur when the
gesture input patterns for two words are too similar, or sometimes completely ambiguous, 
leading to input errors. An example gesture input error is illustrated in Figure
\ref{fig:Example-swipe-collision}. A recent study showed that gesture input has an error rate that 
is about 5-10\% higher compared to touch typing [\cite{octopus}]. With the fast and inherently imprecise
nature of gesture input the prevalence of errors is unavoidable and
the need to correct these errors significantly slows down the rate of text entry.
The QWERTY keyboard in particular is poorly suited as a medium
for swipe input. Characteristics such as the ``u'', ``i'', and
``o'' keys being adjacent lead to numerous gesture ambiguities and
potential input errors. It is clearly not the optimal layout for gesture input. 

The rise of digital keyboard use, first on stylus based keyboards in the 90's and then
on modern touchscreens a decade later, has led to a lot of research and development in 
breaking away from QWERTY to a layout that's statistically more efficient. This work resulted
in various improved layouts for digital stylus keyboards such as the OPTI keyboard [\cite{OPTI}],
the Metropolis and Hooke keyboards [\cite{Metropolis}], and the ATOMIK keyboard [\cite{ATOMIK}].
In addition to statistical efficiency, attempts were also made to improve statistical efficiency while 
simultaneously making the new layout as easy to use for novices as possible [\cite{layoutMatters}].

More recently, a few keyboards have been introduced that improve text input for certain situations on modern smartphones and tablets:
optimizing for the speed of two-thumb text entry on tablets [\cite{oulasvirta2013improving}];
optimizing tap-typing ambiguity (the SWRM keyboard) and simultaneously optimizing single-finger text entry for speed, reduced tap-typing ambiguity, and familiarity with the QWERTY keyboard (the SATH keyboard) [\cite{Dunlop:2012:MPO:2207676.2208659}];
and optimizing the autocorrect feature itself to simultaneously increase the accuracy of word correction and completion [\cite{Xiaojun:Complete}].

Most of the aforementioned work was done specifically for touch typing since that is the most
common form of text input on touchscreen devices. However, the relatively recent rise in popularity of 
gesture typing has led to some interesting new keyboard layouts that were specifically optimized for
improved gesture typing performance. The Square OSK and Hexagon OSK keyboards were optimized
to maximize gesture input speed using Fitt's law [\cite{fastSwiping}]. Various optimizations were also
done by Smith, Bi, and Zhai while maintaining some familiarity with QWERTY by using the same layout 
geometry and only changing the letter placements. This resulted in four new keyboards: the GK-C keyboard,
which is optimized to maximize gesture input clarity; the GK-S keyboard, which is optimized for gesture input
speed; the GK-D keyboard, which was simultaneously optimized for gesture clarity and speed using Pareto
optimization; and the GK-T keyboard, which was simultaneously optimized for gesture clarity, gesture speed, 
and QWERTY similarity [\cite{googleKeyboard}].

Evaluating and comparing various keyboard layouts is a difficult problem given the complexity and variability associated with text entry.
Measuring text entry and error rates from user based trials is typically done to evaluate or directly compare the effectiveness of various keyboards and input methods.
These studies usually require the test subjects to transcribe a set of predefined phrases using a specified input device.
Text entry evaluations of mini-QWERTY keyboards [\cite{miniQWERTY}], chording keyboards [\cite{twiddler}], handwriting recognition systems [\cite{handwriting}], and various gesture input systems [\cite{graffiti}, \cite{joystick}] have all been done using this approach.
The main downside of this approach is the fact that in day-to-day use most users spend very little time transcribing phrases.
The majority of text entry is done by composing original phrases.
Therefore, text entry evaluations from transcription based user studies are not realistic and can introduce unintended biases into the results.
Vertanen and Kristensson showed how these biases can be mitigated by including composition tasks in user trials to complement the standard transcription tasks [\cite{composition}].

Despite the recent work done to improve text entry evaluations with user based studies the metrics used for optimization are typically based on surrogate models of the actual performance characteristic of interest.
For example, the gesture clarity metric used by \cite{googleKeyboard} is correlated with how frequently words are correctly reconstructed but does not measure this directly.
The reason that these approximate measures have been used is that accurately evaluating real keyboard reconstruction error rates would require an immense amount of user input data.
Modern optimization techniques typically evaluate and compare hundreds of thousands of different keyboard layouts, making it completely infeasible to obtain the necessary user data.
The methodology that we propose allows for the direct evaluation of gesture reconstruction error rates, or any other desired metric, by simulating realistic user interactions with a keyboard.
This is similar to the approach used by Fowler et al\mbox{.} when they simulated noisy tap typing input to estimate the effect of language model personalization on word error rate [\cite{Fowler}].

To demonstrate the effectiveness of our methodology we will show how it can be used to find a keyboard layout that minimizes gesture input errors.
This requires accurately modeling gesture input for a given layout, interpreting the input, and quantifying how frequently typical inputs would be misinterpreted.
We employ several different models for gesture input and a dictionary of the most common words in the English language to simulate realistic usage and take into account variations between users.
We also attempt to develop a highly accurate algorithm for recognizing gesture inputs that is not limited to a specific keyboard layout.
It should be noted that although this paper focuses on the error rate performance, the overall methodology can be used to evaluate and compare keyboard layouts based on any performance measure.

Finally, In order to address the problem we designed and built an open source software framework, called Dodona, for exploring different input methods.
This framework is well suited for examining a wide range of possible keyboard designs and input models.
It was built with optimization in mind and has a focus on efficient implementations and extensibility.
The library is freely available on GitHub [\cite{dodona}] and was used to perform the analysis and generate all keyboard related graphics presented here.

\section{\textbf{Modeling Swipe Input}}\label{sec:Modeling-Swipe-Input}

An extremely large dataset of gesture inputs is needed in order to accurately
evaluate the error rate of a given keyboard layout. The only way to 
obtain such a dataset on a reasonable time-scale is to generate
gesture input data based on models of user input. To accomplish
this we developed several models which can take a word and produce
what we refer to as a gesture input vector, a sequential series of $\left(x,\ y,\ t\right)$
points that represent discrete samples along a gesture input pattern. We then
used words that were randomly generated based on their frequency of
use in the English language to feed into these models and generate
realistic sets of input.

\subsection{Input Vectors and Interpolations\label{sub:Input-Vectors-and}}
In general, our input model can produce either a ``random vector''
or a ``perfect vector''. The former is used for realistic, inexact
gesture input while the latter represents the ideal input pattern that
is free from variation. To construct random vectors we begin by drawing
control points for each letter in a given word from a two dimensional
Gaussian distribution that's centered around each corresponding key
on the keyboard. The $x$ and $y$ widths of the Gaussian, in addition
to the correlation in the offsets between subsequent control points,
can be changed as parameters of the input model. We then interpolate
between these control points for each letter to produce a continuous
gesture input as a function of time. This is then sampled at evenly
spaced intervals along the entire interpolation in order to produce an input
vector with a set number of points. Perfect vectors are constructed
in the same way but use the centers of the keys as control points. The
idea that their exists a unique perfect vector for each word in the lexicon was first 
introduced by Kristensson and Zhai in their seminal paper about the SHARK$^2$ 
text input system for stylus keyboards [\cite{SHARK2}]. In their work they 
refer to perfect vectors as \textit{sokgraphs}.

\begin{figure}[tp]
\begin{centering}
\subfloat[Random Input Vectors]{
	\includegraphics[scale=0.51]{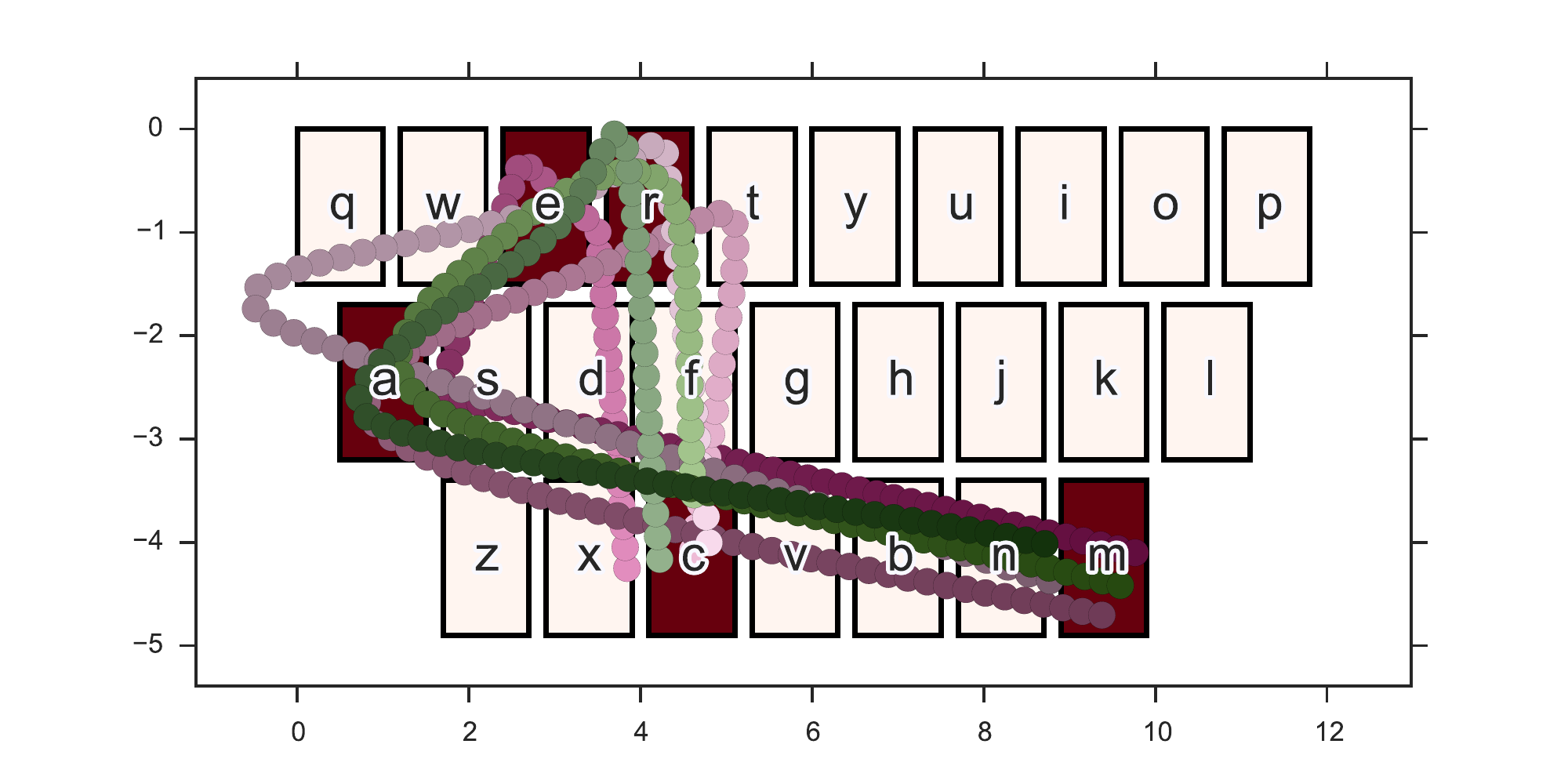}
} \\
\subfloat[Perfect Input Vectors]{
	\includegraphics[scale=0.5]{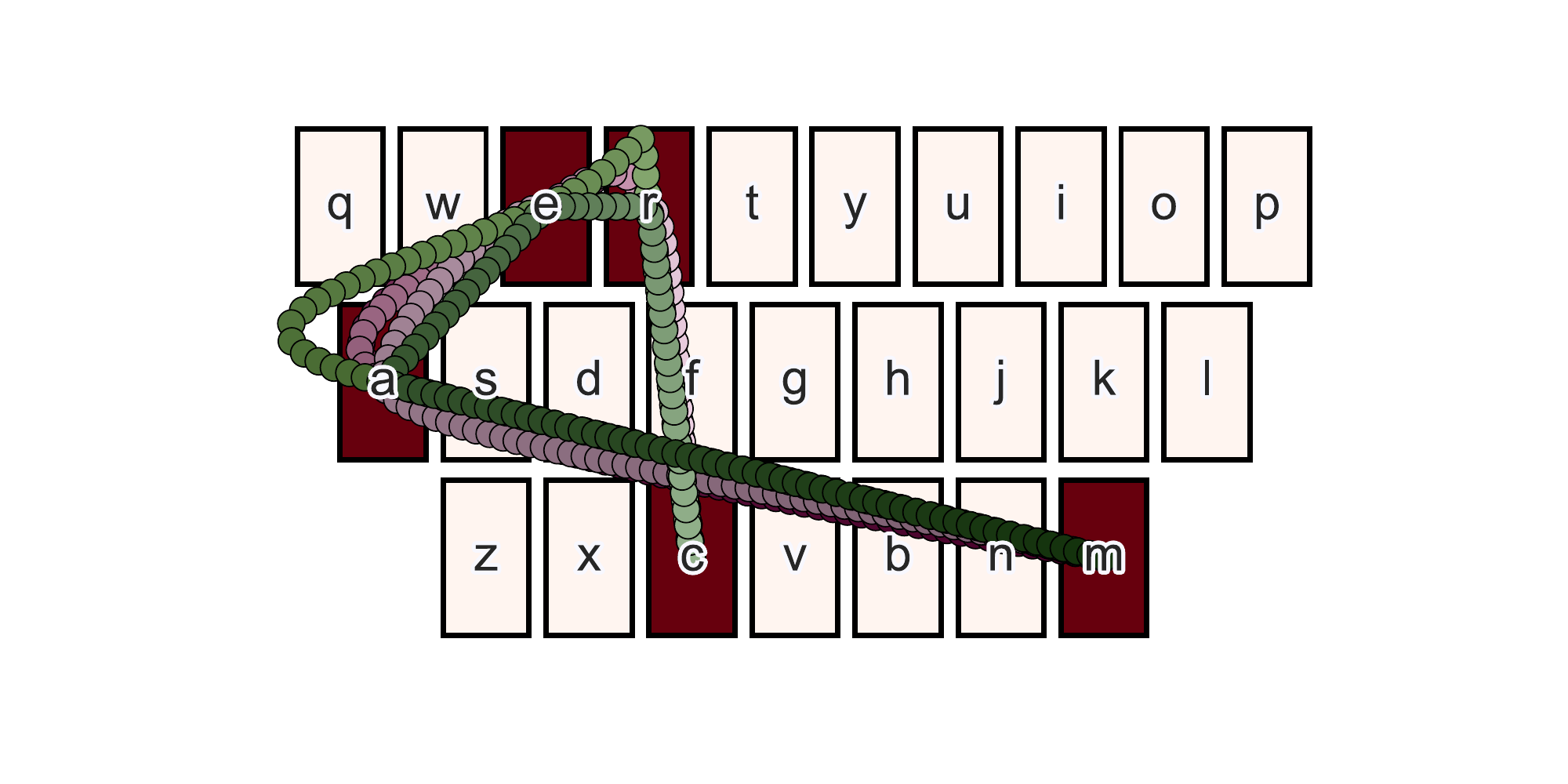}
}
\par\end{centering}
\protect\caption{Five random swipe input vectors for the word ``cream'', each one
using a different type of interpolation and randomly generated control
points.\label{fig:Five-different-random}}
\end{figure}

We chose to implement a variety of different interpolations to account
for the variations in individual gesture input style. We settled on five
different interpolation techniques: a straight-line spatial interpolation,
a natural cubic spline, a cubic Hermite spline [\cite{Splines}], a monotonic cubic
Hermite spline [\cite{MonotonicSplines}], and a modified natural cubic spline where the first
and last segments are required to be straight lines.

Using randomly generated control points with various interpolation
techniques allows us to capture a large range of input possibilities.
This is demonstrated in Figure \ref{fig:Five-different-random}, which
shows five different possible swipe patterns corresponding to the
word ``cream''. Each pattern was constructed with a different interpolation
and random variations of the control points.

\subsection{Lexicon\label{sub:Word-Corpus}}
\begin{figure}[tp!]
\begin{centering}
\includegraphics[scale=0.5]{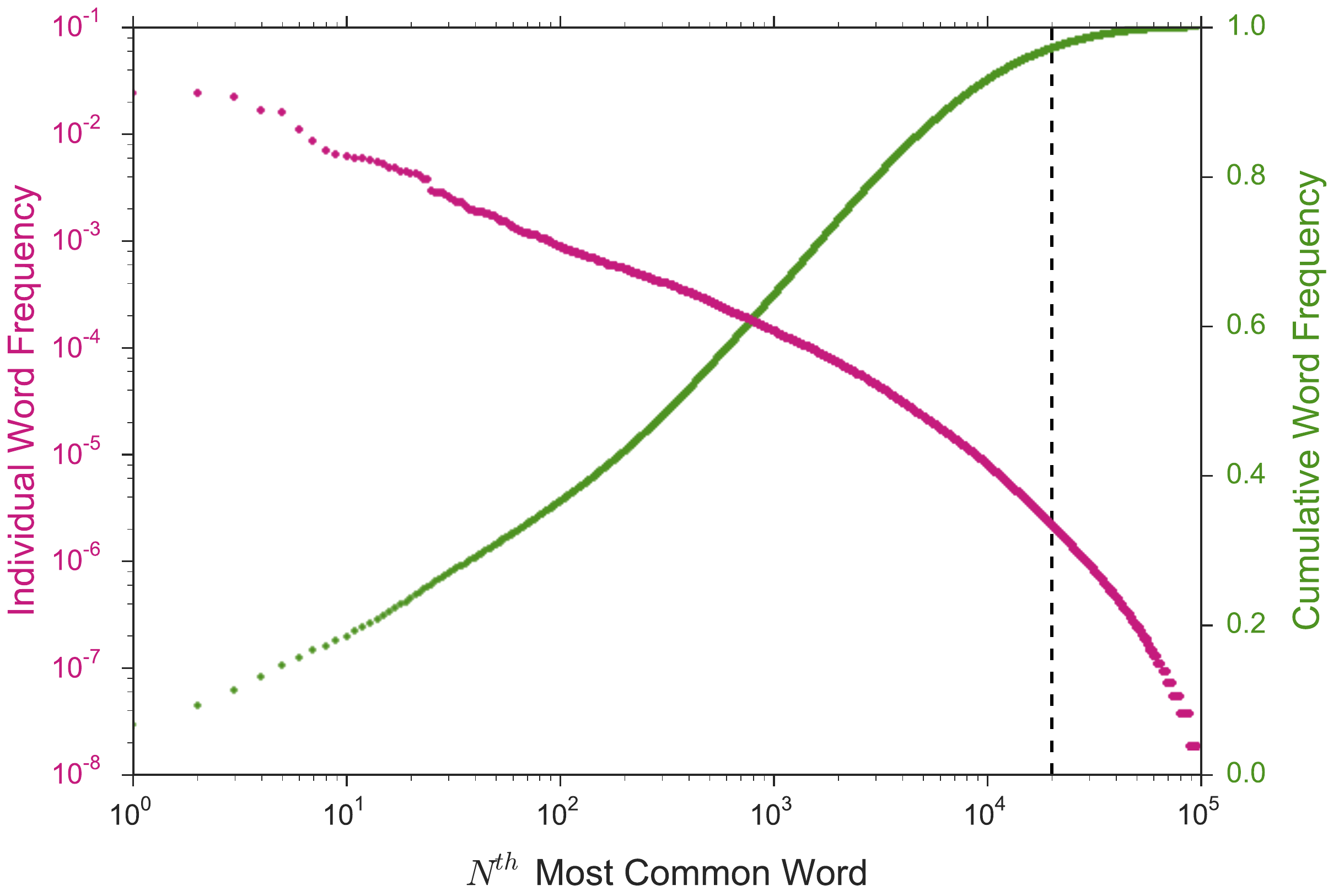}
\par\end{centering}
\smallskip{}
\protect\caption{Individual word frequencies in the lexicon (magenta) and the cumulative
distribution (green). The vertical black dashed line indicates the
$20,000$ word marker, which is where we cut off for inclusion in
the analysis dictionary.\label{fig:Individual-word-frequencies}}
\end{figure}

We used the Google Web Trillion Word Corpus as compiled by Peter Norvig
to provide the dictionary for our input model [\cite{Norvig:Misc}].
This dictionary contains approximately 300,000 of the most commonly
used words in the English language and their frequencies. However,
more than half of the entries are misspelled words and abbreviations.
Our final dictionary only contained words from the Google Web Trillion
Word Corpus that also occurred in the Official Scrabble Dictionary,
the Most Common Boys/Girls Names, or WinEdt\textquoteright s US Dictionary
[\cite{Scrabble:Ward,WinEdit:Wordlist}]. The result was a dictionary
containing 95,881 English words and their frequencies. The individual
word frequencies (magenta) and the associated cumulative distribution
(green) are shown in Figure \ref{fig:Individual-word-frequencies}.
In order to reduce the computational needs associated with using a
dictionary this large we elected to only use the top $20,000$ words
in the optimization procedures described later. Even though this is
only $20.9\%$ of the words contained in the dictionary it accounts
for $97.2\%$ of the total word occurrences. Furthermore, the average
vocabulary size of a college educated adult in the U.S. is $17,200$
words with a range extending from about $13,200$ to $20,700$ words,
which is consistent with the size of the dictionary used in this analysis [\cite{GOULDEN1990}].

\section{\textbf{Gesture Clarity and Recognition Error Rate}}\label{sec:gcMethods}
In order to evaluate the gesture clarity of a given keyboard layout, a quantitative metric
must be defined. In the recent paper by Smith, Bi, and Zhai they define gesture clarity as
the average distance between a vector representation of each word and its nearest neighbor on a given keyboard layout [\cite{googleKeyboard}].
The vector used to represent each word corresponds to what they call its \textit{ideal trace} (identical to \textit{perfect vectors} defined in this paper).
This definition is naturally related to how effective a keyboard will be since keyboards with smaller distances between nearest neighbors will tend to have more frequent reconstruction errors.
However, there are a number of important factors that it does not take into account:
more than one neighboring word can be a source of recognition errors, there are threshold distances beyond which the impact on gesture clarity is negligible, and there are countless subtle interplays between the specific layout of a keyboard and the way that users input imperfect gestures.
Therefore, a different procedure for computing something akin to gesture clarity is required if we want to take these effects into account and more accurately reflect the frequency of words being successfully reconstructed.

For this reason we decided to use the \textit{gesture recognition error rate} for a given keyboard layout and lexicon as our metric.
Ideally, this would be measured experimentally but this is time consuming, expensive, and essentially impossible when the goal is to evaluate a large number of keyboards as would be done in an optimization.
In the absence of real user testing, modeling and simulating how users input gestures allows for an approximate evaluation of the actual error rate.
The error rate, $e$, for a given keyboard layout can be approximated by generating $N$ random input vectors from random words in the lexicon.
The generated input vectors can then be reconstructed using a realistic algorithm and checked against the words that they were meant to represent.
If $n$ input vectors were reconstructed as the wrong word then the recognition error rate is simply $e = \frac{n}{N}$.
This quantity can very roughly be thought of as relating to the gesture clarity, $c$, according to $c\sim1-e$, though this relationship is just a heuristic. As mentioned in the previous paragraph, there are subtle nuances that contribute to gesture input errors that can affect the error rate of a given keyboard layout but not its gesture clarity.

\subsection{Gesture Input Recognition}\label{sub:Distance-Measure}
Since our error rate metric depends on how accurately we can recognize gesture inputs on a given keyboard layout we need a gesture input recognition algorithm that can accurately recognize input vectors on any keyboard layout.
Using our input model we can define gesture input recognition as the ability correctly match a random input
vector with the word that generated it\footnote{On real keyboards it is the ability to match a gesture input with the word
that the user intended to enter. If our input model is accurate than these two definitions are essentially identical.}.
This is a difficult problem to solve because as a user passes their finger/stylus over the keyboard they typically pass over
many more characters than those required to spell the desired word.
This means that we must be able to pick out a particular word from
an ambiguous input. If you look at the example gesture input pattern for the
word ``whole'' in Figure \ref{fig:Example-swipe-collision}, it
is easy to see how even differentiating between two words can be a
challenge.

\subsubsection{Euclidean Distance}
Our first approach to this problem was to simply take the Euclidean distance between two input vectors. 
This requires each input vector to be normalized so that they each have the same number of interpolation points, which are equally spaced along the interpolation.
Implementing the Euclidean distance approach is then straightforward and given by the equation,

\begin{equation}
D=\sqrt{\sum_{i=1}^{n_{ip}}\left[(x_{1,i}-x_{2,i})^{2}+(y_{1,i}-y_{2,i})^{2}\right]},\label{eq:cartDist}
\end{equation}
where $n_{ip}$ is the total number of interpolation points in the gesture input vector and $x_{1,i}$ is the $x$-component of the $i$th interpolation point in the first of the two input vectors being compared. 
This is very similar to the \textit{proportional shape matching} channel used in the SHARK$^2$ writing system [\cite{SHARK2}] and in the gesture clarity
metric used by \cite{googleKeyboard}. 

Although this method can correctly identify when two gesture inputs match exactly, it could also return a large distance between two input vectors that are qualitatively very similar.
For example, a user may start near the bottom of the key for the first letter of the word and end up shifting the entire gesture input pattern below the centers of the subsequent keys.
This input pattern could pass over all of the correct keys but still result in a large Euclidean distance when compared to the perfect vector for the word. 

The shortcomings of this approach made it clear that we were not utilizing all of the useful information contained in the input vectors.
If a poor distance measure were to cause misidentifications that would not happen in practice then this could introduce significant biases during the optimization procedure, resulting in a keyboard that is not optimal for real use.
Kristensson and Zhai accounted for this in the SHARK$^2$ writing system by incorporating language and location information in addition to proportional shape matching in their recognition algorithm.
Similarly, In order to reduce the impact of these systematic effects, we needed to identify additional features that would improve our gesture input recognition.

\subsubsection{Feature Set}
Our first step was to uncouple the $x$ and $y$ coordinates and treat
them individually. Given the anisotropic nature of most keyboards, the
relative importance of biases between the two spatial dimensions is
not clear a priori. Therefore, we decided that the first two features
should be the Euclidean distance (Eq. \ref{eq:cartDist})
between two input vectors for the $x$ and $y$ components individually,
\begin{equation}
D_{x}=\sqrt{\sum_{i=1}^{n_{ip}}(x_{1,i}-x_{2,i})^{2}}\textrm{{\ \ and\ \ }}D_{y}=\sqrt{{\sum_{i=1}^{n_{ip}}(y_{1,i}-y_{2,i})^{2}}}.\label{eq:DxDy}
\end{equation}

\begin{figure}[t]
\begin{centering}
\includegraphics[scale=0.22]{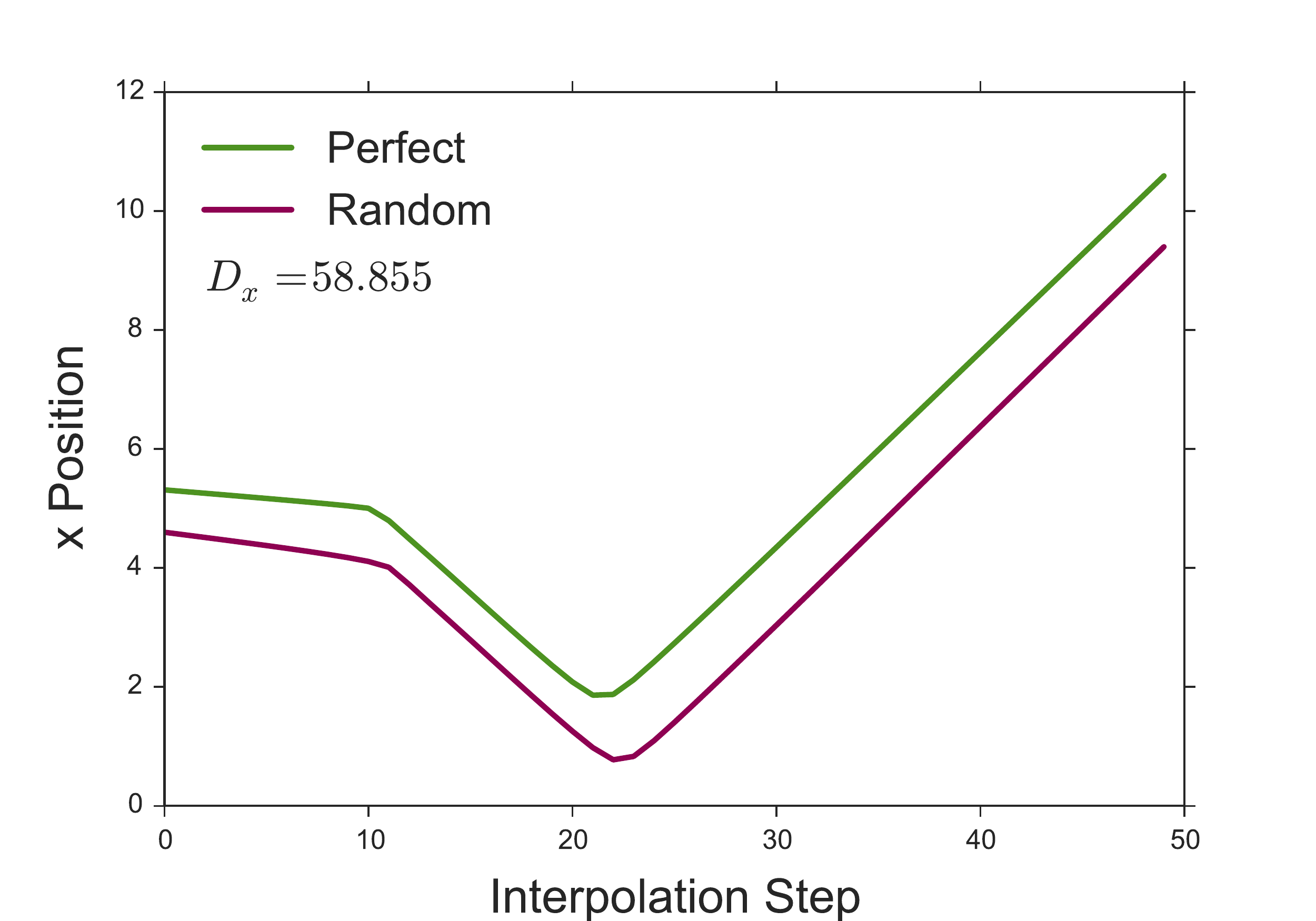}\includegraphics[scale=0.22]{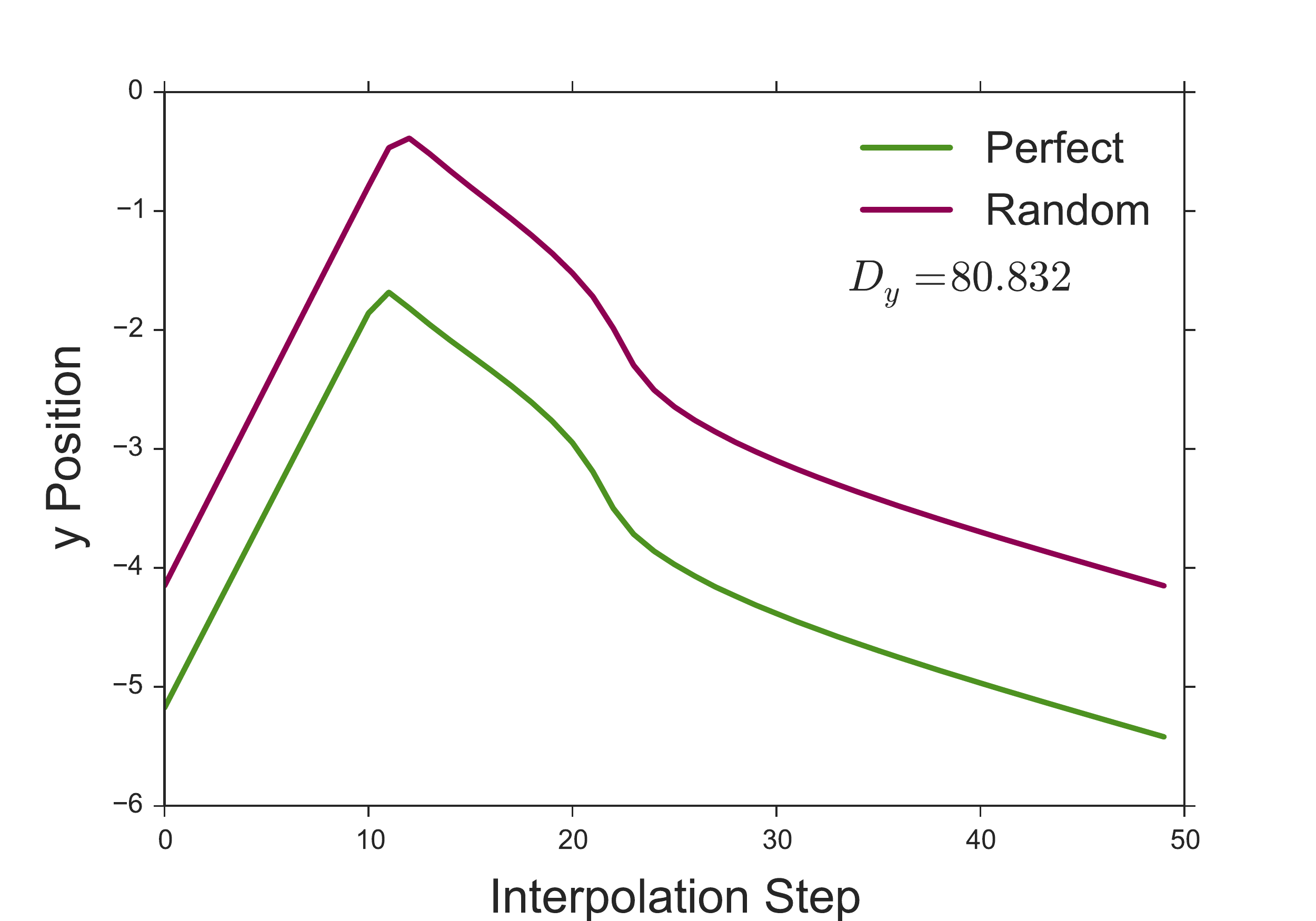}
\par\end{centering}
\begin{centering}
\includegraphics[scale=0.22]{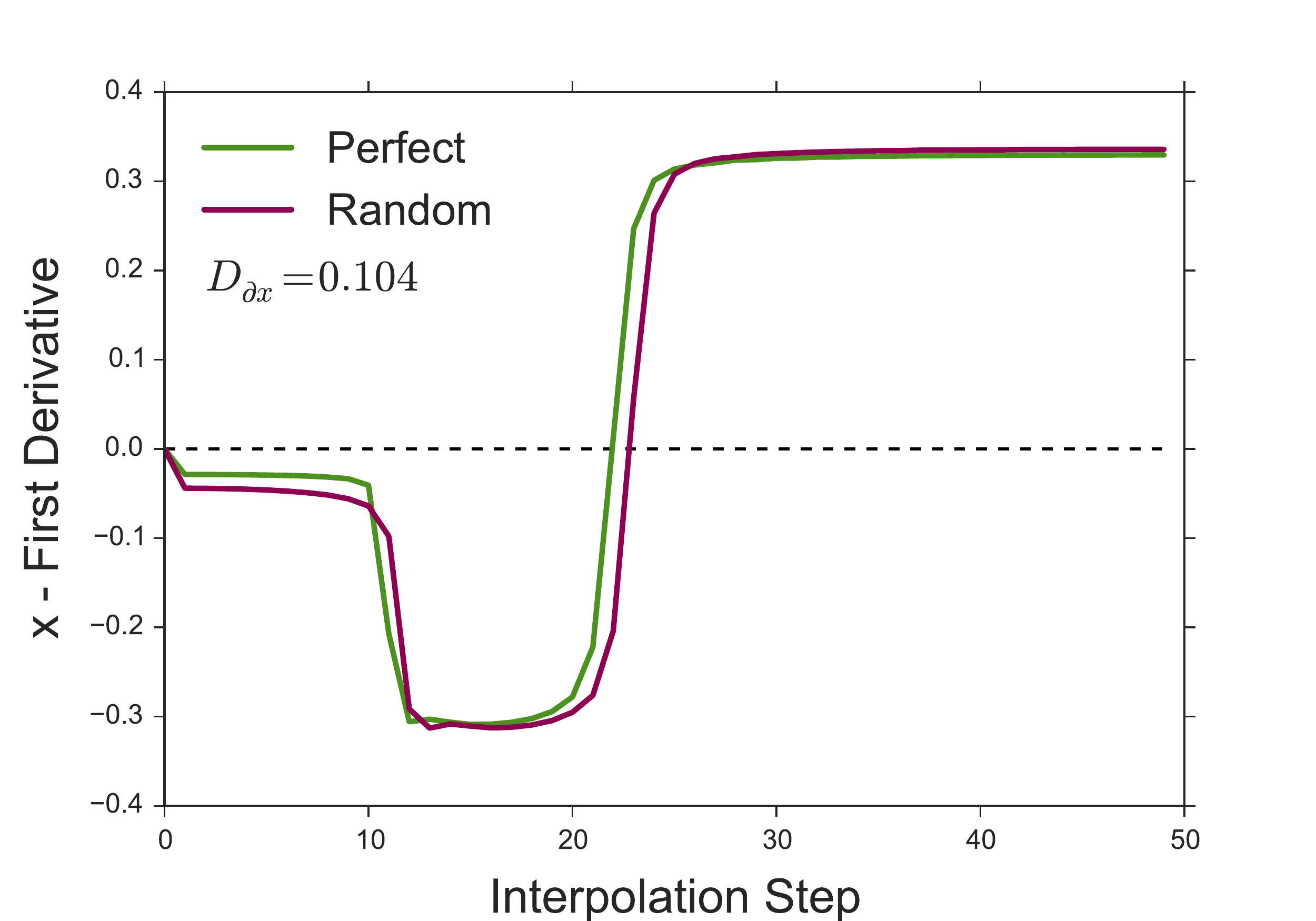}\includegraphics[scale=0.22]{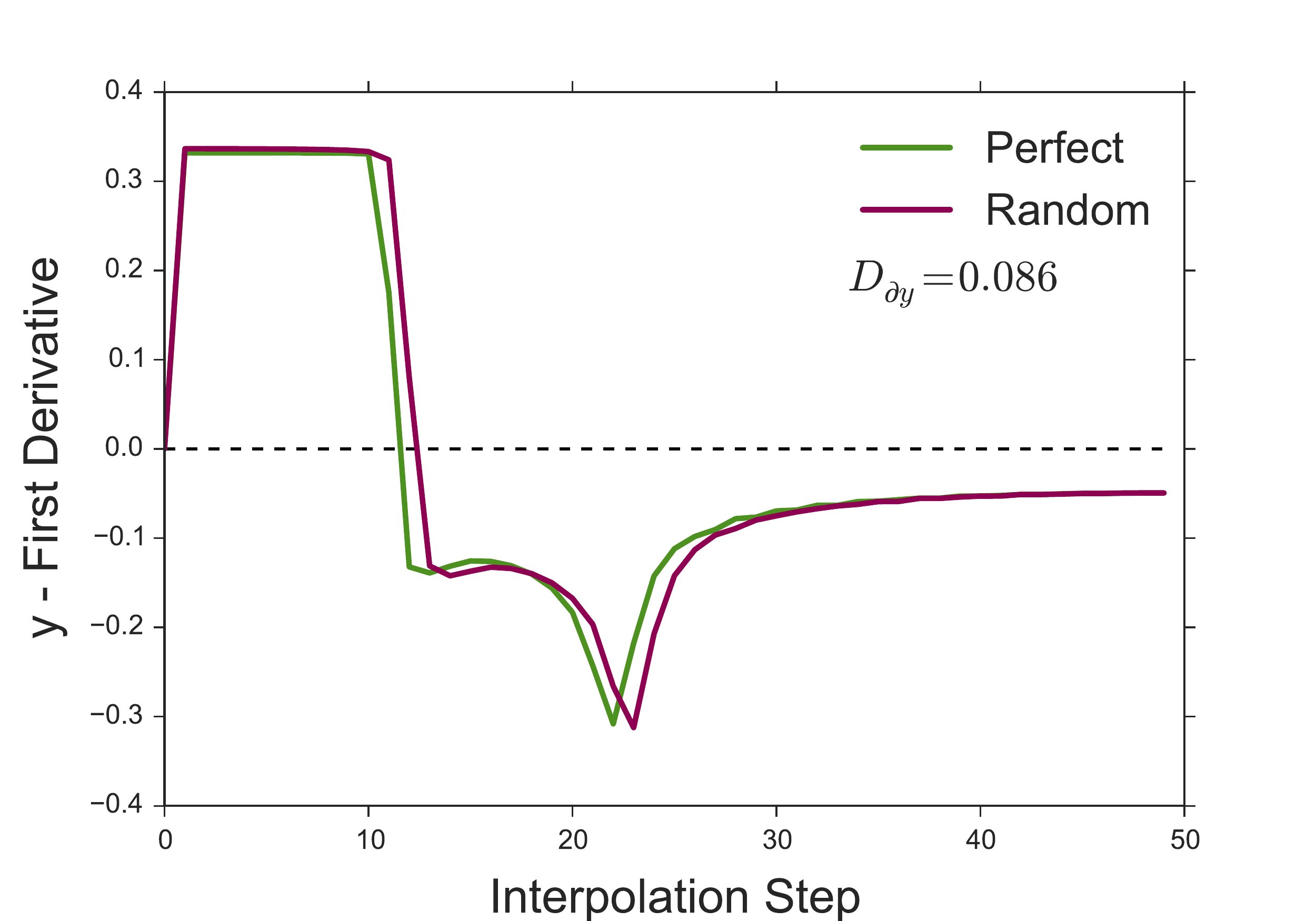}
\par\end{centering}
\begin{centering}
\includegraphics[scale=0.22]{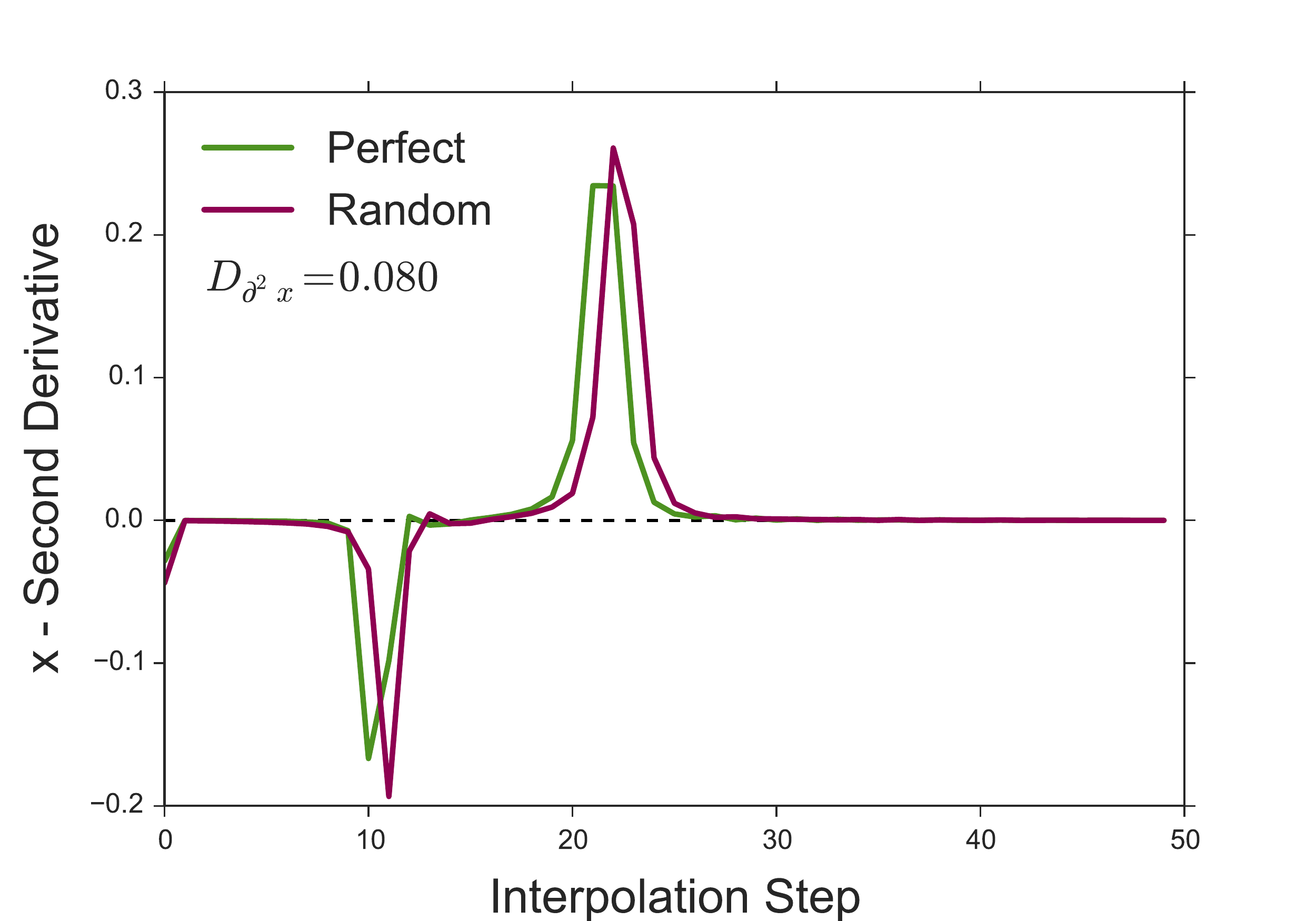}\includegraphics[scale=0.22]{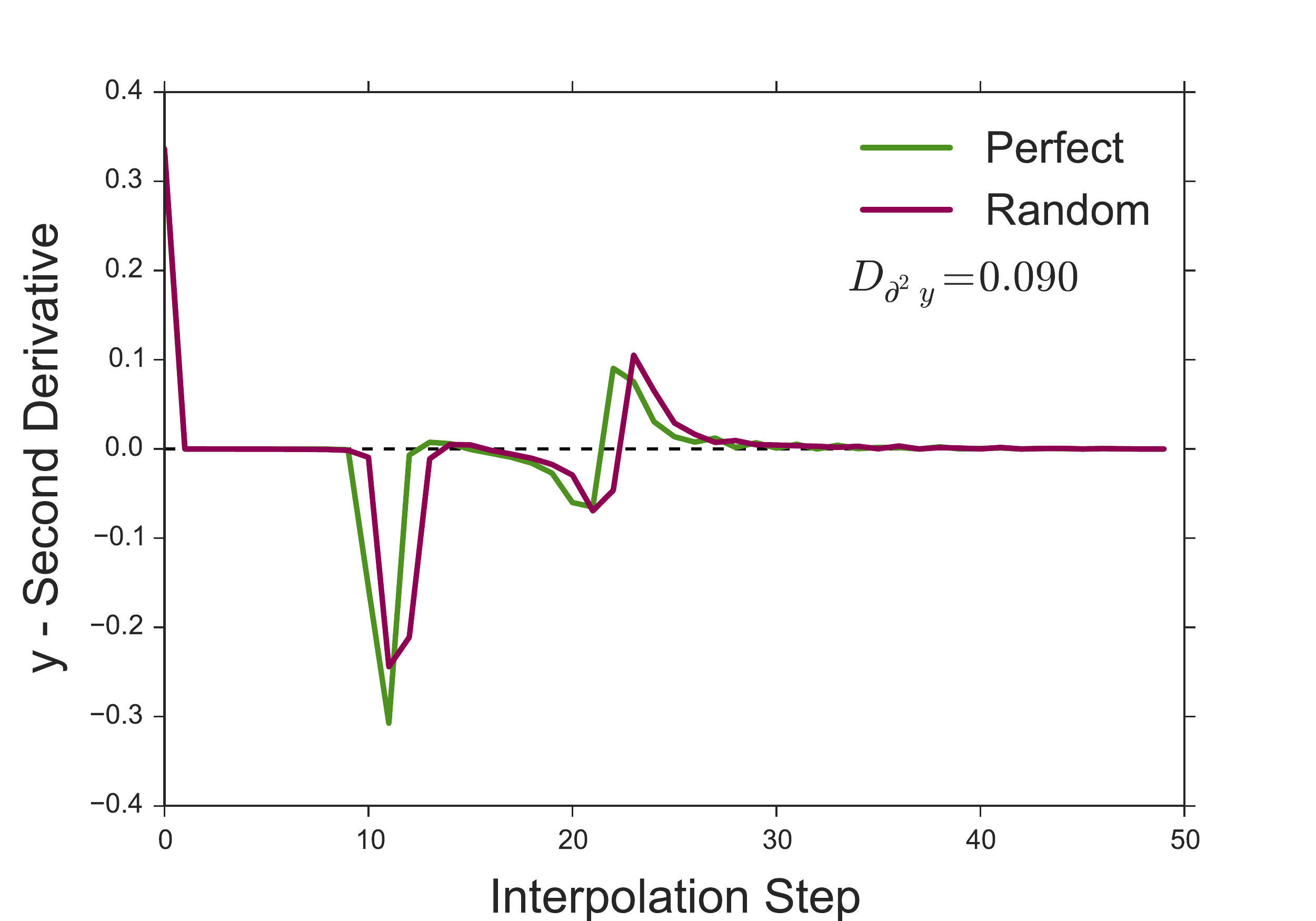}
\par\end{centering}
\medskip{}
\protect\caption{A closer look at a random input vector (magenta) and a perfect vector
(green) for the word ``cream''. The x-component (top), its derivative
(middle), and its second derivative (bottom) are shown in the left
column while the same information for the y-coordinate is shown on
the right. The corresponding values that would go into the feature
set $(D_{x},D_{y},...)$ are also shown in each plot.\label{fig:A-closer-look}}
\end{figure}

In order to address the issue with offset input vectors, translational
symmetry needed to be taken into account. To do this we decided to
look at the time derivatives of the input vectors with respect to
their $x$ and $y$ coordinates. Since the input vectors are sets
of sequential, discrete points we can easily estimate the derivative
at each point. We can then construct a distance measure by taking
the Euclidian distance between the time derivatives of two swipe patterns
at each point. The equation for the derivative distance in the $x$-dimension
is given by:

\begin{equation}
D_{\partial x}=\sqrt{{\sum_{i=1}^{n_{ip}-1}\left[(x_{1,i+1}-x_{1,i})-(x_{2,i+1}-x_{2,i})\right]^{2}}},\label{eq:dDist}
\end{equation}
where $x_{1}$ and $x_{2}$ are the $x$-components of the two input
vectors being compared. We assume a constant input velocity for the
models considered here so we've implicitly rescaled the time coordinated
such that $t_{i+1}-t_{i}=1$ to simplify the equations.

We also wanted a distance measure that would be more sensitive to
the positions of sharp turns in a gesture input pattern. This led us to include
the distance between the second derivatives of the input vectors
in a similar fashion to the first derivatives (Eq. \ref{eq:dDist}).
The quantity $D_{\partial^{2}x}^{2}+D_{\partial^{2}y}^{2}$ is rotationally
invariant as well so we can see how these might help allow for more
leniency in situations where there might be some variation in the
orientation of a touchscreen device relative to a users hand.

The utility of these features in regards to correctly identifying
gesture input is apparent when you take a closer look at the differences
between a random input vector and a perfect vector for a given word.
The $x$ and $y$ values as a function of time for a random input
vector and a perfect vector for the word ``cream'', as well as their
first and second derivatives, are shown in Figure \ref{fig:A-closer-look}.
This example illustrates how the first and second derivatives can
be useful for finding the correct match even when the swipe pattern
is shifted significantly from the center of the appropriate keys.

Two additional distinguishing features of each gesture input pattern are the
start and end positions. These points carry information about the
overall position of an input vector while being less sensitive to
the shape than $D_{x}$ and $D_{y}$. The addition of this information to the Euclidean distance was
shown by \cite{SHARK2} to reduce the number of perfect vector ambiguities by ~63\% 
in their 20,000 word lexicon. Consequently, the distance
between the $x$ and $y$ components of the first and last points
of the input vectors were included in the feature set. This gives us four additional
features to add to the six previously discussed.

Finally, we realized that the length of each gesture input pattern can be
a very distinguishing characteristic, leading us to include the difference
in the length of each gesture input pattern as the final addition to our feature
set, giving us a total of eleven features which are related to the
difference between two gesture input patterns. However, in order avoid repeatedly
calculating square roots we decided to put the squared value of each
distance in the feature set. To summarize, the set contains: the squared
Euclidean distance between the $x$ and $y$ components, the squared distance
between the $x$ and $y$ components of the first derivatives, the squared 
distance between the $x$ and $y$ components of the second derivatives,
the squared distance between the $x$ and $y$ components of the first
point, the squared distance between the $x$ and $y$ components of
the last point, and the difference in the squared length of the two
gesture input patterns being compared.

\subsubsection{Artificial Neural Network Classifier}
The last step in creating our desired distance measure was figuring
out a way to combine the eleven elements in our feature set to a single
output representing the ``distance'' between two gesture input vectors.
Despite the intuitive basis of our feature set, the relationship of
each element to the overall performance of a classifier is highly
non-trivial given the complexity of gesture input vectors. Fortunately,
this is a problem that is well suited for a neural network based solution.

To build a classifier using our feature set, we created a deep, fully-connected
artificial neural network with eleven input nodes; one for each of
the variables in the previously discussed feature set. The network
architecture consists of three hidden layers with 11 nodes each and
a fourth hidden layer with only two nodes. The activation function
for each hidden node and the output node is an Elliot function,

\begin{equation}
\begin{centering}
f_{E}(x)=\frac{0.5sx}{1+s|x|}+0.5,
\end{centering}
\end{equation}
where $s$ is the steepness and is set to 0.5 for each layer. This
is a computationally efficient sigmoid function that asymptotes
slower than a standard sigmoid. This was necessary since we employed
the RPROP algorithm [\cite{rprop}] to train the network, which is
susceptible to the flat-spot problem when using steep sigmoid activation
functions. The artificial neural network used in the analysis was
implemented using the Fast Artificial Neural Network software library
(FANN) [\cite{nissen03}] and is available in the repository as a FANN binary
file and as a text file listing the weights of every connection in the deep neural network [\cite{dodona}].

\begin{figure}[tbh]
\begin{centering}
\includegraphics[scale=0.5]{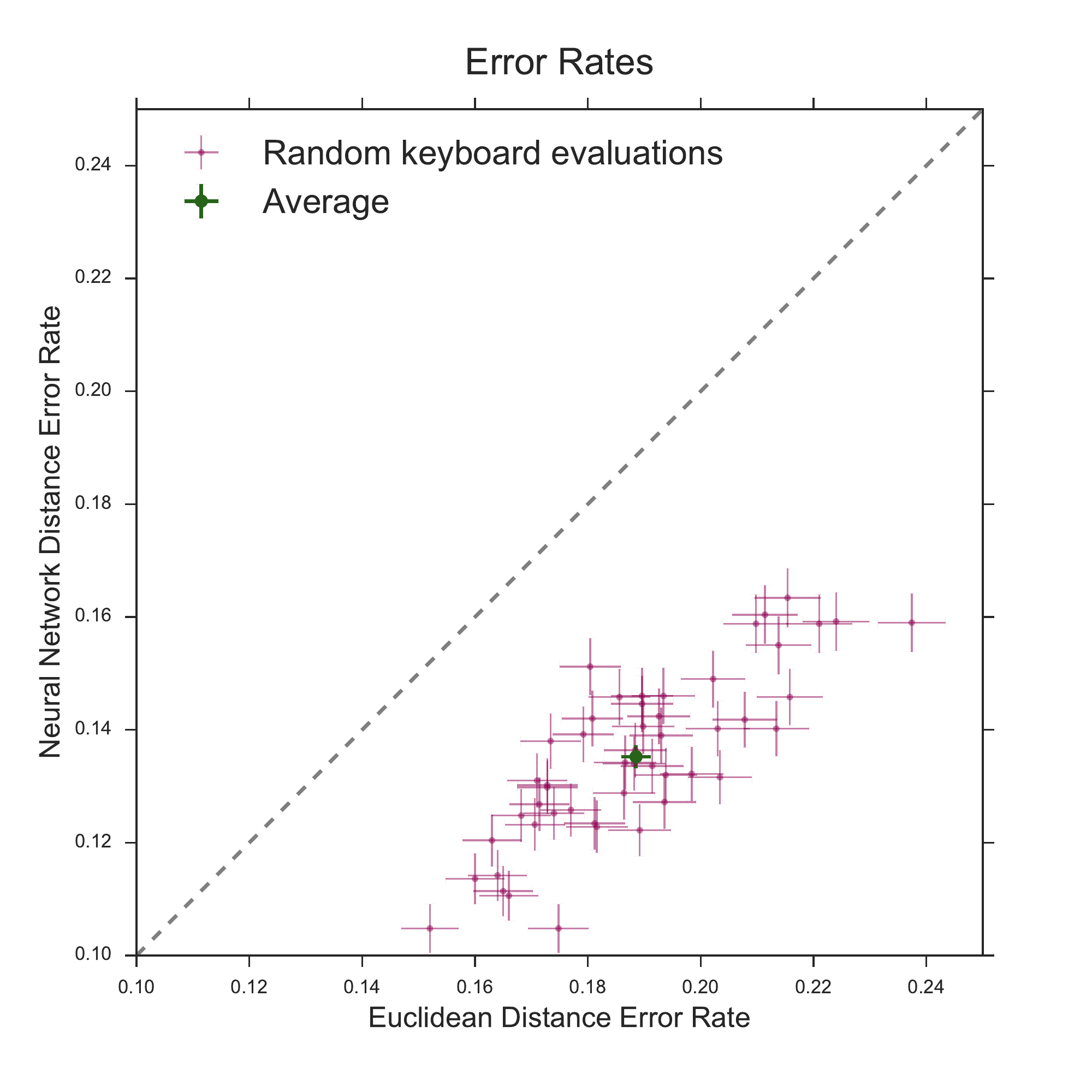}
\par\end{centering}
\smallskip{}
\protect\caption{A comparison of the gesture recognition error rate for two 
different gesture recognition algorithms. Each point in the scatter plot shows the
error rate of a given (random) keyboard layout as calculated using the Euclidean distance
measure ($x$-axis) and the neural network distance measure ($y$-axis). The dashed line
is just $y=x$ and is only meant to help the comparison. The average error rate for each 
method is represented by the bold green dot.\label{fig:The-performance-of-NN}}
\end{figure}

The neural network was trained on a dataset of $500,000$ pairs of
random input vectors and perfect vectors. The random input vectors
were constructed with each of the different interpolation models used
($\sim$100,000 for each model) but the perfect vectors
were restricted to be linear interpolations. This was done to make
the algorithm as realistic as possible since a practical gesture recognition algorithm
would not be able to make any assumptions about a user's input style.
The training set was divided up so that $30\%$ of the input pairs
corresponded to the same word, $20\%$ corresponded to completely
random words, and the remaining $50\%$ corresponded to different
but similar words. The exact definition of ``similar words'' is
given in the next section. 

The performance of the neural network recognition algorithm can be seen in Figure \ref{fig:The-performance-of-NN}.
This plot shows a comparison between the neural network method and the Euclidean distance method.
In this study, 50 random keyboards layouts were created and for each layout 5,000 random gesture input vectors were generated.
The input vectors were then matched to a word using the two methods. 
The fraction of attempts that each method got wrong is shown as the error rate (as discussed in much more detail in the next section).
It is easy to see that for each keyboard layout the neural network recognition algorithm outperformed the Euclidean distance algorithm. 
The average error rate using the neural network algorithm on random  keyboard layouts is $13.5\% \pm 0.2\%$ compared to $18.9\% \pm 0.3\%$ for the Euclidean distance measure, which is a reduction of $28.5\%$ in the gesture input recognition error rate.

\subsection{Monte Carlo Error Rate Evaluation\label{sub:Error-Rate-Evaluation}}
With a more accurate gesture recognition algorithm we can confidently evaluate 
a given keyboard's gesture recognition error rate. The general approach is to use a Monte Carlo based algorithm to determine
the error rate. This technique can be described as follows: a random
word is chosen from the lexicon with a probability proportional
to its frequency of occurrence. A random gesture input vector is then
generated for this word based on a given input model. 
The gesture recognition algorithm then determines the word that is the
best match for that specific random input vector. If the selected word matches the original
word then the match is considered a success. This process is repeated
$N$ times so that the error rate is given by the ratio of successful
matches to the total number of attempts. Due to the statistical nature
of this technique there will be an uncertainty in each measurement.
As with most efficiency calculations, the uncertainty is given by
the variance of a binomial distribution scaled by $\frac{1}{\sqrt{N}}$.

Although effective, this method is very computationally intensive.
A reasonable optimization procedure will require around $5,000$ matching
attempts in each efficiency calculation to reduce the effects of statistical
fluctuations (specifically, this produces error rate calculations with statistical uncertainties of $\sim$0.7\%).
Each matching attempt requires a comparison for every
word in the lexicon, which contains $20,000$ words, so every efficiency
determination will require $100,000,000$ distance measure calculations.
Since the goal is to use the error rate calculation in an optimization
procedure, increasing the total time by several orders of magnitude,
another approach was needed.

\subsubsection{Radix Tree Pruning}
Consider the case where a user is trying to input the word ``pot''
on a standard QWERTY keyboard. Clearly words such as ``cash'' and
``rules'' are not going to be calculated as the best match by the
distance measure because they have dramatically different gesture input patterns.
Therefore, there is no need to spend time comparing the perfect vectors
of these words as part of the error rate calculation. The error rate
calculation can be made much faster without sacrificing much accuracy
by comparing only to the perfect vectors of more plausible candidate
words such as ``pit", ``put", ``lit'', etc. The difficulty lies only in determining which words
are plausible candidates.

\begin{figure}[t]
\begin{centering}
\includegraphics[scale=0.9]{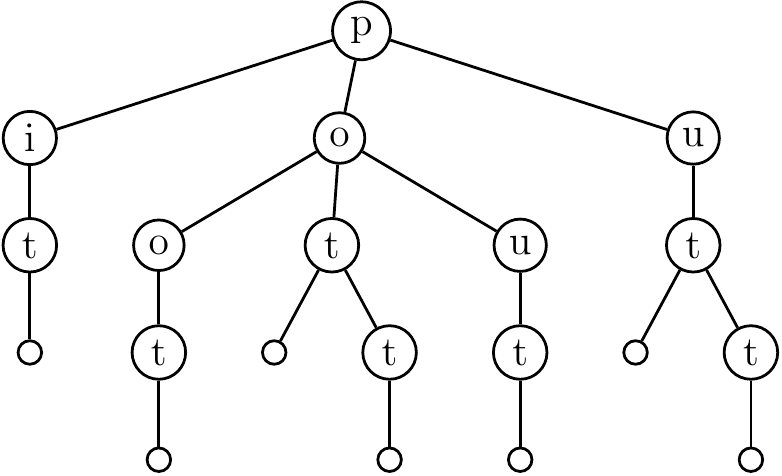}
\par\end{centering}
\medskip{}
\protect\caption{A Radix tree showing some of the branches consistent with an input
vector for the word \textquotedblleft pot\textquotedblright{} and
the potential string form \textquotedblleft poiuyt.\textquotedblright \label{fig:A-Radix-tree}}
\end{figure}

To determine what words should be included in the candidate list we
created what we call the ``string form'' for each gesture input vector.
The string form is just the sequential list of letters that the gesture input
pattern traversed. For example, if a user inputs the word ``pot''
the corresponding string form might be ``poiuyt''. If we were implementing
an algorithm for determining candidates given a fixed keyboard layout
we could first generate a large number of random input vectors for
every word in the dictionary. We could then build a lookup table where
each observed string form corresponds to a list of all words that
had been observed to ever produce that string form.

That approach is unfortunately not possible when optimizing the keyboard
layout because the lookup table would need to be rebuilt for every
new keyboard. Instead, we generate a large number of random vectors
for the word that would be the correct match and find the string form
for each of those. We then allow any words that are contained within
any of these string forms to be potential candidates. This would not
be possible if we didn't know the intended word but it results in
a superset of the candidates we would find using the precomputed lookup
table given that a sufficient number of random vectors are used. 

The words that are consistent with each string form are determined
by recursively searching a radix tree representation of all of the
words in the dictionary as shown in Figure \ref{fig:A-Radix-tree}.
In this example we would start by looking at the radix tree that begins
with the letter ``p''. Then we would look to see if this tree has
a child node corresponding to the letter ``p''; repeated letters
are not always distinguishable in a swipe pattern so we have to always
explicitly check for the possibility. Since there is no word that
begins with ``pp'' we then move on to the next letter in the string
form, ``o'', and look for a child node corresponding to this letter.
The search is then done recursively for the sub-tree beginning on
the child node corresponding to ``o'' with the string form ``oiuyt''.
The search continues in this recursive manner, traversing all of the
branches of the subtree that are consistent with the string form and
returning any leaf that is found as a candidate word. This will effectively
find all candidate words beginning with ``po'' that could potentially
be contained in the string form. Once this subtree is exhausted we
move on to the next letter of the original string form, ``i'', and
recursively search the subtree corresponding to this letter with the
string form ``iuyt'' for candidate words beginning with ``pi''.
This process continues until the final subtree corresponding to the
letter ``t'' is searched, thus finding all candidate words contained
in the string form.

\begin{figure}[t]
\begin{centering}
\includegraphics[scale=0.43]{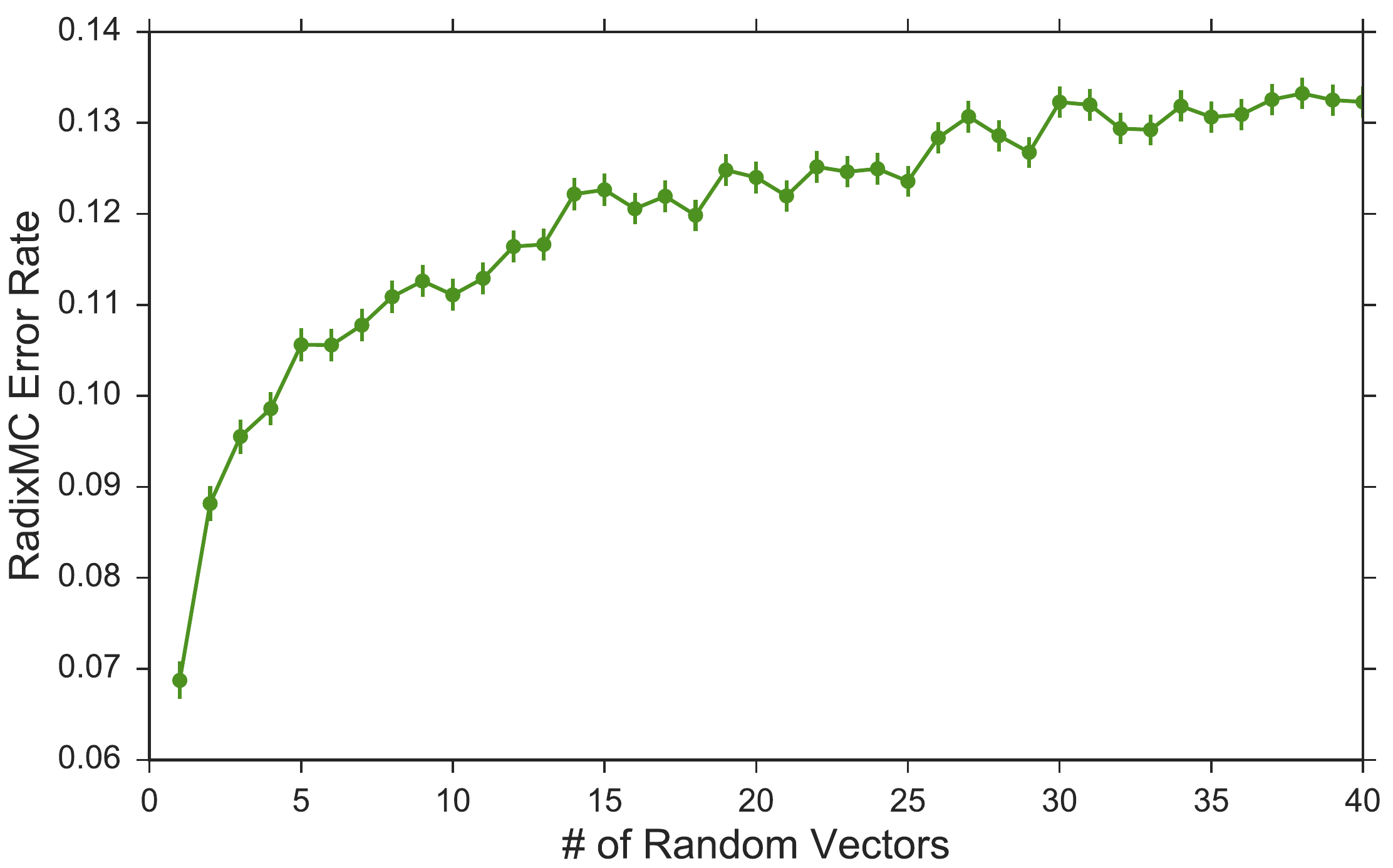}
\par\end{centering}
\smallskip{}
\protect\caption{The error rate calculated using the RadixMC method as a function of 
the number of random vectors used in the radix tree pruning step.\label{fig:radix_vs_nvectors}}
\end{figure}

This approach, which we call \textit{radix tree pruning} (abbreviated as "RadixMC" when combined with the standard Monte Carlo algorithm), reduces the number of comparisons to make for each input vector and subsequently speeds up the calculation significantly.
The time required scales roughly linearly with the number of random vectors which are used to find candidates so some balance between efficiency and accuracy is required.
In order to determine a suitable threshold we calculated the error for a given random keyboard as a function of the number of random vectors used in the radix tree pruning as shown in Figure \ref{fig:radix_vs_nvectors}.
The flattening out of the error rate is expected since the words that are most similar are typically found in the earlier iterations.

We chose to use 20 random vectors in the pruning step since they allow for the vast majority of nearest neighbors to be found and can be generated in a reasonable amount of time.
We conducted a small study where we calculated the error rate while varying the number of random vectors from one to 25 for the QWERTY keyboard and four random keyboard layouts.
The results showed that the relative error rate of the random keyboards to QWERTY remained roughly constant across the entire range of the number of random vectors.
The relationship between the error rate and the number of input vectors is largely independent of the keyboard layout, which means the relative error rates remain approximately unchanged.
When performing an optimization, where only relative error rates are important, this effect becomes largely negligible.

\begin{figure}[t]
\begin{centering}
\includegraphics[scale=0.5]{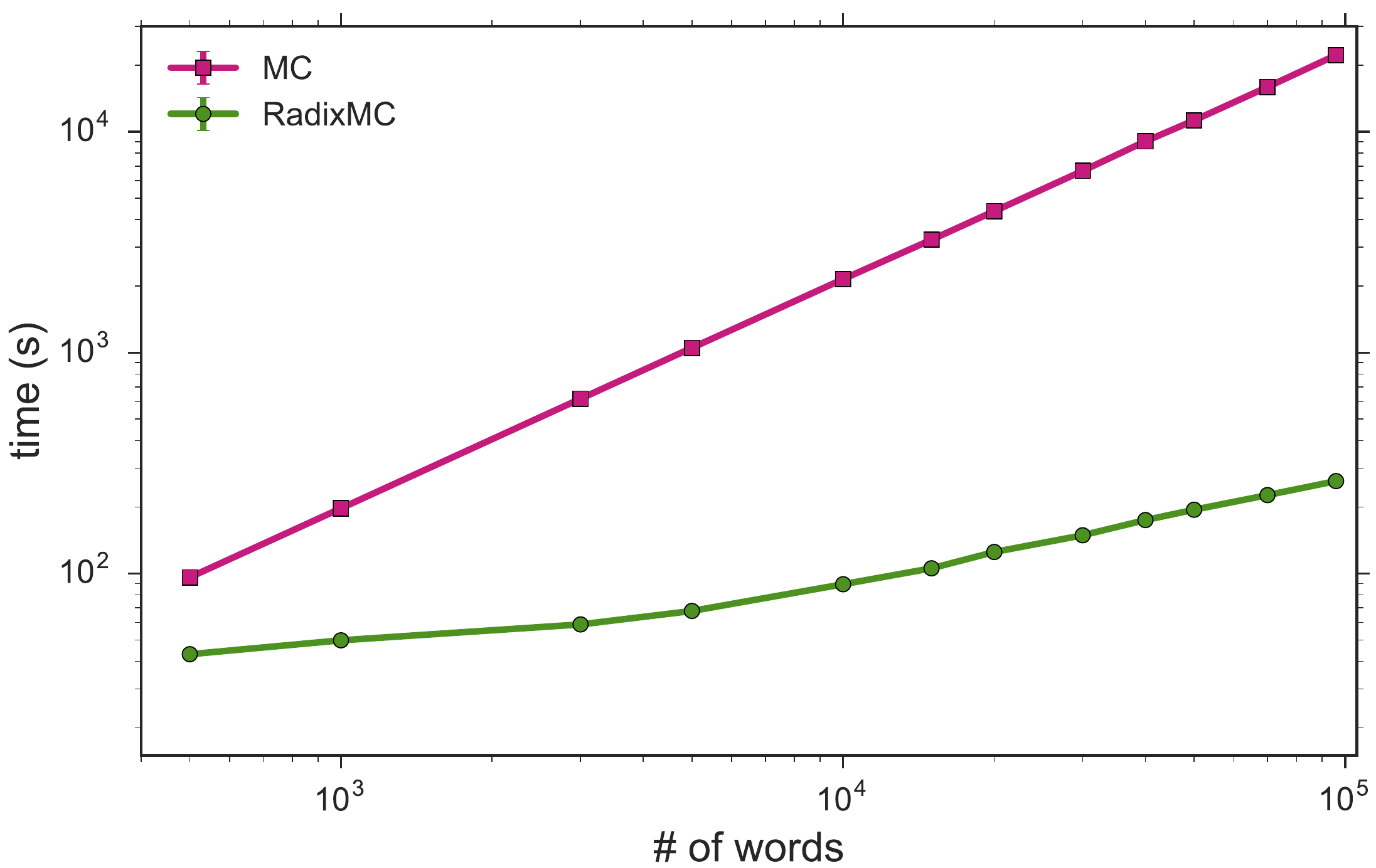}
\par\end{centering}
\smallskip{}
\protect\caption{The CPU time required to do an error rate calculation using the basic
Monte Carlo approach (magenta squares) and the RadixMC approach (green circles) as
a function of the number of words in the dictionary.\label{fig:The-cpu-time}}
\end{figure}

When used in the error rate calculation this algorithm outperforms the standard Monte Carlo approach in terms of computational efficiency by two orders of magnitude when the full dictionary is used, as seen in Figure \ref{fig:The-cpu-time}.
It is also obvious that the radix tree based Monte Carlo algorithm scales much more favorably with the size of the dictionary.

A similar problem was faced by Kristensson and Zhai during the development of SHARK$^2$.
However, they opted for a different filtering technique, which they called \textit{template pruning}.
This required that the start and end positions of the input vector and a perfect vector be less than a given threshold in order to be considered for a potential match [\cite{SHARK2}].

Both of these approaches have the same goal: to drastically reduce the number of comparisons necessary to determine the intended input word.
The main difference lies in the trade-off between the computational complexity of the filtering and the number of comparisons made during the matching stage.
While computationally very efficient, template pruning has the consequence of passing many words to the comparison step that would be filtered using the radix tree approach.

Using Kristensson's and Zhai's method with a pruning threshold large enough to allow for consistent word reconstruction, we were able to prune the list of comparison words by 95.08\% on average.
The radix tree pruning technique with 20 random vectors was able to reduce the necessary list of comparison words by 99.92\%.
By removing so many additional candidate words that have nearby starting and ending keys but dissimilar input vectors, we were able to significantly reduce the time required for the combined filtering and matching stages.

\section{\textbf{Results}}\label{sec:Results}

\subsection{Optimizing Keyboards to Minimize the Error Rate \label{sub:Keyboard-Optimization}}
In order to demonstrate the utility of our methodology and the Dodona open source framework we optimized permutations of a standard keyboard layout to the minimize gesture recognition error rate.
The specific optimization algorithm we employed begins with generating
a random keyboard layout and calculating its error rate. At each
successive iteration a new keyboard is created by swapping $n$ pairs
of keys in the previous keyboard. The error rate of the new keyboard
is calculated for each interpolation method and averaged. The average
error rate is then compared to the previous keyboard. If the error
rate has decreased then the new keyboard is kept and the previous
one is discarded, otherwise the new keyboard gets discarded. This
process is repeated $N$ times where the number of key swaps, $n$,
is repeatedly decreased by one at set intervals. This results in successive
keyboards differing by only one swap at the end of the optimization
procedure. For our final analysis we ran 256 separate optimizations,
each running through 200 iterations, and starting with $n=6$. 

\begin{figure}[t]
\begin{centering}
\includegraphics[scale=0.5]{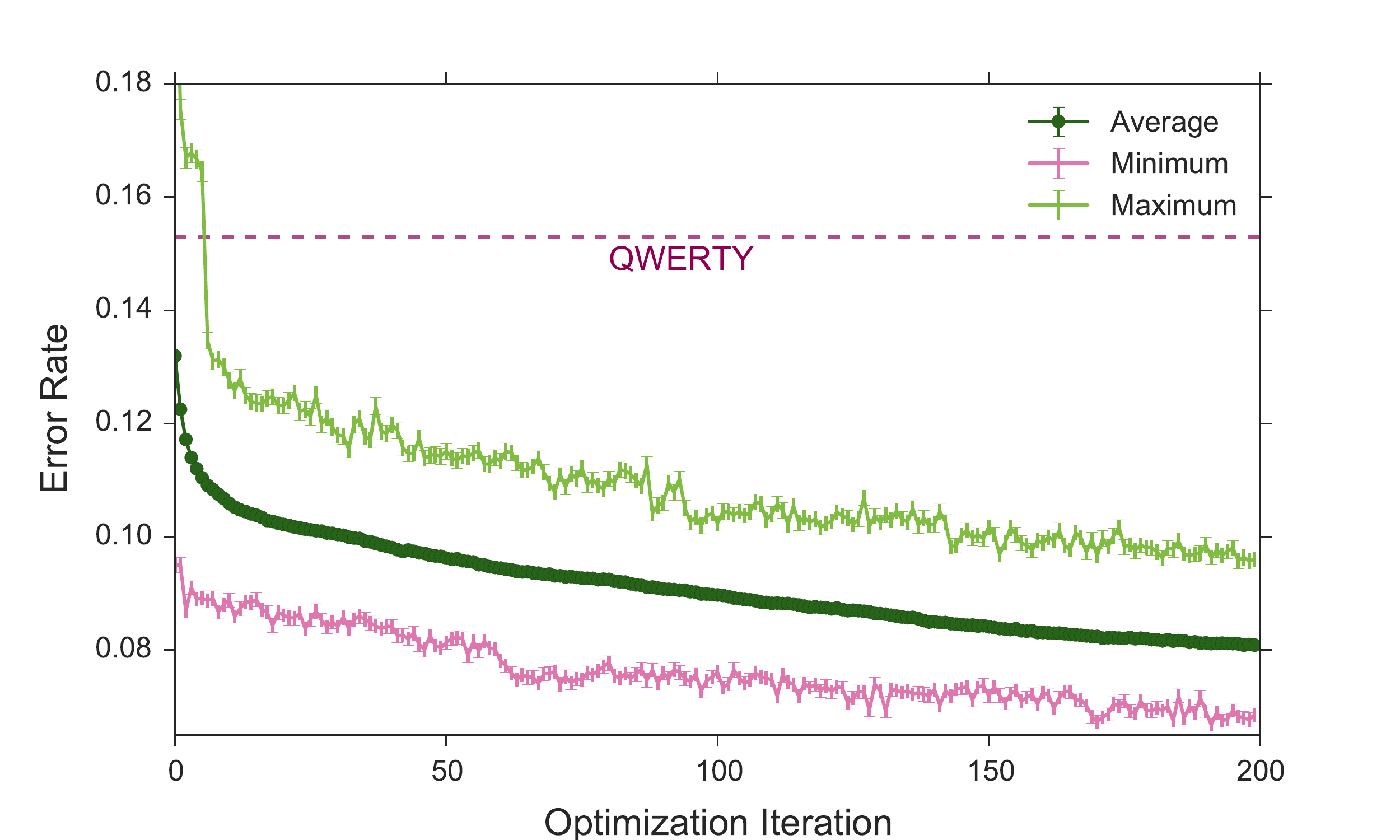}
\par\end{centering}
\smallskip{}
\protect\caption{The average error rate (dark green) as a function of optimization
step resulting from the procedure outlined in the methods section.
Also shown are the minimum (pink) and maximum (light green) error
rates calculated at each optimization step. For comparison, the error
rate for the QWERTY keyboard (as determined by our model) is shown
by the dashed magenta line.\label{fig:The-Average-error}}
\end{figure}

The average error rate for all of the input models at each step in the optimization
procedure is shown in Figure \ref{fig:The-Average-error}. The minimum
and maximum error rate at each step and the error rate of the QWERTY
keyboard are also shown in the figure. Interestingly, we see that
the QWERTY keyboard error rate of $15.3 \pm 0.04$\% is less efficient for gesture
input than the average randomly generated keyboard layout (error rate:
$\sim$13\%). However, the optimization procedure quickly finds keyboard 
layouts with even lower error rates. After two hundred iterations the average error 
rate found in each trial is approximately 8.1\%. This represents an improvement in the
error rate of 47\% over the QWERTY keyboard. 

\begin{figure}[tbh]
\begin{centering}
\includegraphics[scale=0.65]{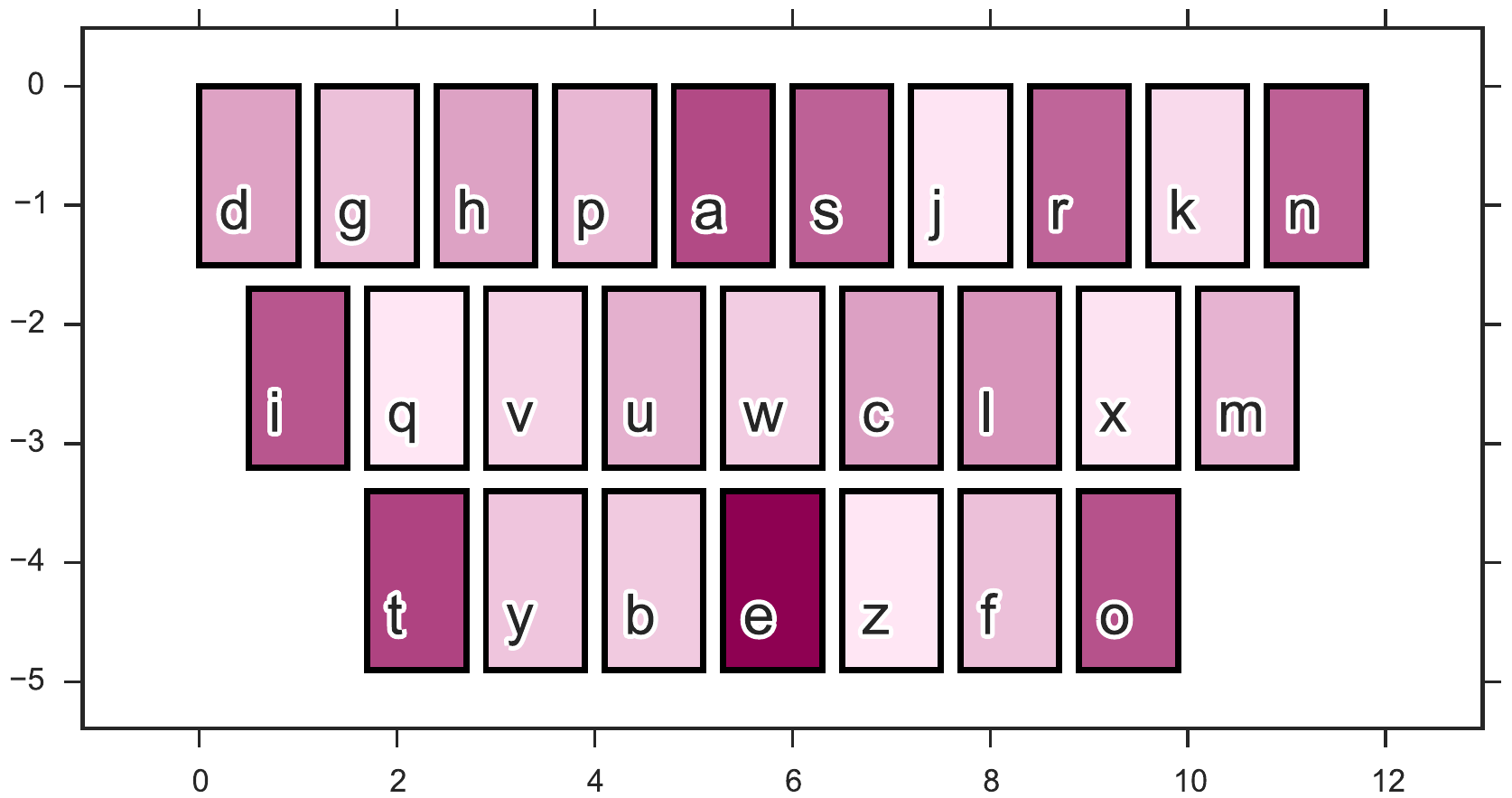}
\par\end{centering}
\medskip{}
\protect\caption{The most optimal keyboard found by optimizing keyboard layouts to
minimize the error rate for swipe input. The shade of magenta on each
key indicates how frequently that key is used in the lexicon.
Dark magenta corresponds to the most frequent usage and white/light
magenta the least frequent. \label{fig:The-most-optimal}}
\end{figure}

The optimal keyboard for gesture input clarity found in the analysis is shown in Figure \ref{fig:The-most-optimal}
and we will refer to it by its first four letters: \textit{DGHP}.
The keys are colored to represent their relative frequency in the
lexicon. This keyboard is found to have an error rate of $7.67 \pm 0.06\%$,
which is a $50.1 \pm 0.4\%$ improvement compared to the QWERTY keyboard.
This is higher than the minimum shown in Figure \ref{fig:The-Average-error}
because of the limited resolution of the error rate measurement used
in the optimization procedure. The quoted error rate of the optimal
keyboard was determined by a final high precision measurement.

It is important to note that the values of the error rates computed by
our method depend heavily on the parameters of the input model. Thus,
the error rates themselves hold little general meaning. Instead, it
is more meaningful to speak of the relative change in error rate compared
to the standard QWERTY keyboard. We have found the ratios of the error
rates produced from different keyboards to be largely independent
of the input model parameters. This permits us to state, in general,
whether the keyboard layout resulting from the above procedure is
more efficient than the QWERTY keyboard for swipe input and quantify
the relative improvement.

One major feature of the DGHP keyboard is the appearance
of the most frequently used keys along the edges. The QWERTY keyboard
has many of the most frequently used keys near its center, which results
in large number of gesture input errors. In this gesture input optimized keyboard
the keys at the center of the layout are less frequently used. This
makes sense because having the most frequently used keys at the edges
will decrease the probability that the user passes over them without
intending to . By removing the most common letters from appearing arbitrarily
in gesture input patterns we naturally reduce the number of errors.
However, there are more subtle characteristics of the keyboard that
arise due to the way words are structured in the English language.
For example, the letters ``i'', ``o'', and ``u'' are no longer
clustered together, which eliminates the ambiguity between words like
``pout'', ``pit'', and ``put''. In addition, another notable
feature is the separation of the letters ``s'', ``f'', and ``t''.
This helps to distinguish between the words ``is'', ``if'', and
``it'', which are very common in the English language. It's interesting
to try and understand some of the reasons why the keyboard has such
a low error rate but, in reality, it is a finely tuned balance that
depends on the structure and frequency of every word used in the analysis.

\begin{figure}[t]
\begin{centering}
\includegraphics[scale=0.65]{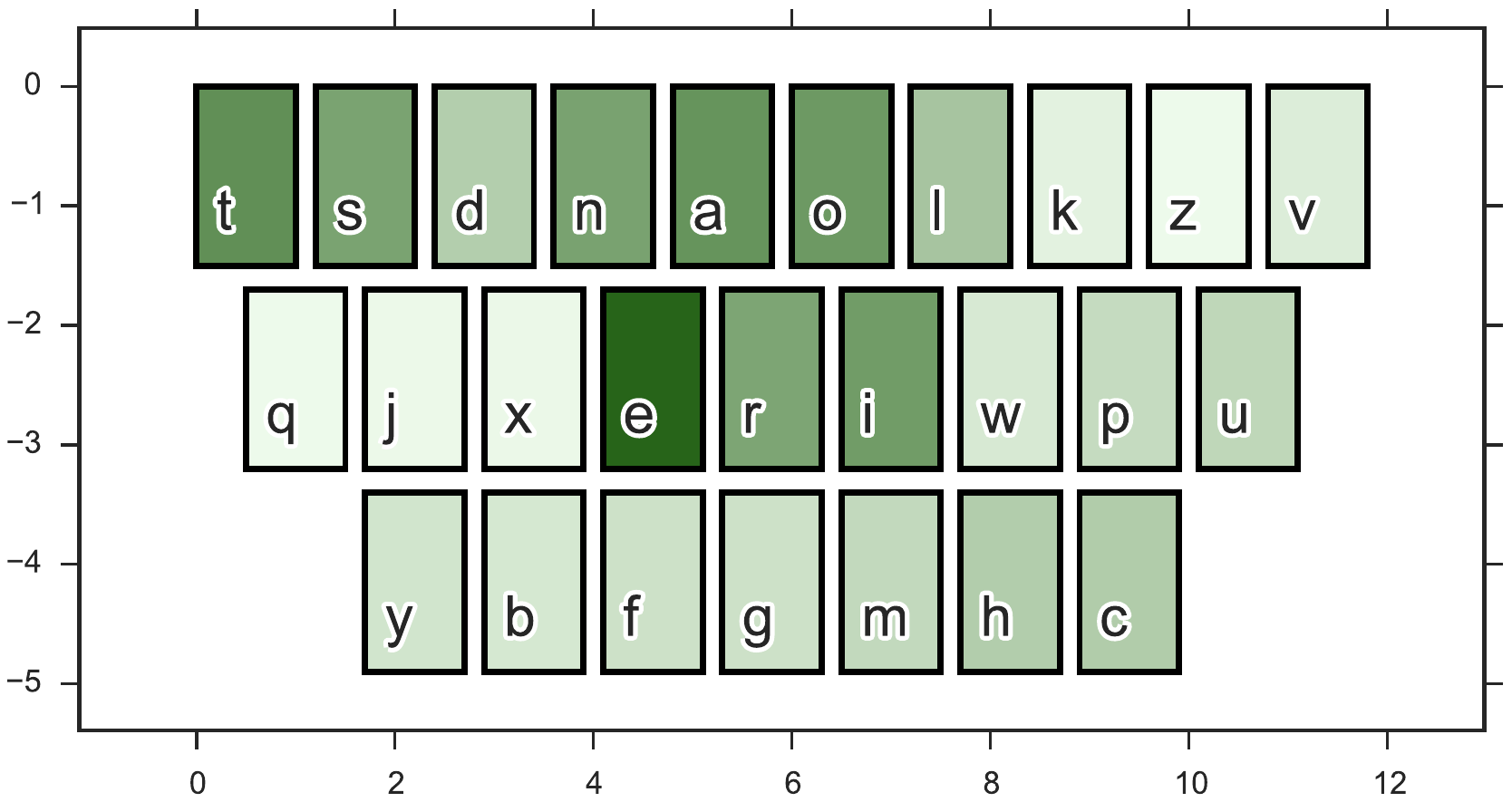}
\par\end{centering}
\medskip{}
\protect\caption{The least optimal keyboard for minimizing the error rate of swipe
input. The shade of green on each key indicates how frequently that
key is used in the lexicon. Dark green corresponds to the most
frequent usage and white/light green the least frequent.\label{fig:The-least-optimal}}
\end{figure}

Out of curiosity we also decided to see what would happen if we optimized
a keyboard to maximize swipe errors. We ran five similar optimization
procedures through 100 iterations to find the least optimal keyboard
layout for swipe input. The worst keyboard we could find is shown
in Figure \ref{fig:The-least-optimal} and has an error rate of 27.2\%,
which is about 78\% worse than the QWERTY keyboard. In this keyboard
the most frequently used keys are all clustered together, making swipe
patterns more ambiguous and resulting in more swipe errors.

\subsection{Evaluating Existing Virtual Keyboard Layouts}
Besides running optimization procedures, evaluating and comparing existing keyboard layouts is another 
important reason to have a robust framework for evaluating keyboards. We evaluated the gesture recognition error rate
for a number of existing virtual keyboard layouts, similarly to what was done by \cite{fastSwiping} when he compared the 
tapping speed for almost\footnote{Radial keyboards like Cirrin and Quickwriting have too much dead space between letters
for this type of input model to be effective and were therefore excluded. In addition, the Montgomery keyboard contains duplicate
letters which, unfortunately, is not currently handled by our software framework and was likewise excluded} 
the same set of keyboard layouts. In addition to the ones presented in his paper we also included the four
keyboards from \cite{googleKeyboard} (GK-C, GK-S, GK-D, GK-T) and the optimal keyboard found here that was presented
in the previous section, DGHP.
The results are displayed in Table \ref{table:keyboard_evaluations}.
The error rate and statistical uncertainty associated with each calculation are displayed for each keyboard. 

The error rates were calculated using 20, 000 Monte Carlo iterations, 20 random vectors in the radix tree pruning step, and random input vectors containing 50 spatially (linearly) interpolated points. As has been noted previously, the absolute value of the error rates is not meaningful unless training data is used to constrain the model. When only the permutations of keys on a single keyboard geometry are considered, however, then the relative error rates of the different layouts can be directly compared because the degree of randomness in the input model scales all of the error rates in a very similar way. The comparison of error rates becomes far more subtle when considering entirely different keyboard geometries, particularly if they included keys of different shapes (e.g. squares and hexagons).

When working with multiple keyboard geometries, there are several different possible approaches for adjusting the input model uncertainty and scaling the relative sizes of the keyboards. In the results shown here, each keyboard was scaled such that the total area of its keys would be the same for all keyboards. We then set the input model uncertainties, $\sigma_x$ and $\sigma_y$, to be the same for all keyboards with values equal to $0.25\sqrt{A_{key}}$, where $A_{key}$ is the area of a single key. It is quite possible, if not likely, that the size and shape of different keys may have an affect on the magnitude of $\sigma_x$ and $\sigma_y$ in real usage. As an alternative approach, we also considered the case where $\sigma_x$ and $\sigma_y$ scale directly with the horizontal and vertical extent of each key, respectively. This resulted in a 10-25\% increase in the error rates of the hexagonal keyboards relative to those with square keys and so this should be taken into account when considering the results in Table \ref{table:keyboard_evaluations}.

\begin{table}[!th]
\centering
\begin{tabular}{l*{2}{c}}
\textit{Keyboard Layout}	& \textit{Gesture Recognition Error Rate} & \textit{Error Rate Uncertainty} \\
\hline
GK-C				& $7.850\%$ & $\pm0.126\%$ \\
SWRM				& $9.003\%$ & $\pm0.148\%$ \\
GK-D				& $9.457\%$ & $\pm0.128\%$ \\
GK-T					& $9.535\%$ & $\pm0.132\%$ \\
Hexagon QWERTY		& $10.074\%$ & $\pm0.134\%$ \\
ATOMIK				& $10.606\%$ & $\pm0.131\%$ \\
Metropolis II			& $10.725\%$ & $\pm0.129\%$ \\
Square Alphabetic		& $10.994\%$ & $\pm0.144\%$ \\
GAG I				& $11.170\%$ & $\pm0.128\%$ \\
Wide Alphabetic		& $11.220\%$ & $\pm0.140\%$ \\
Metropolis I			& $11.344\%$ & $\pm0.137\%$ \\
SATH-Trapezoid		& $11.361\%$ & $\pm0.133\%$ \\
SATH-Rectangle		& $11.531\%$ & $\pm0.152\%$ \\
Chubon				& $11.915\%$ & $\pm0.133\%$ \\
Getschow et al.			& $12.395\%$ & $\pm0.138\%$ \\
OPTI I				& $12.547\%$ & $\pm0.144\%$ \\
Square OSK			& $12.592\%$ & $\pm0.135\%$ \\
Square ATOMIK		& $12.681\%$ & $\pm0.142\%$ \\
Hexagon OSK			& $12.901\%$ & $\pm0.132\%$ \\
QWERTY				& $13.078\%$ & $\pm0.154\%$ \\
Fitaly				& $13.833\%$ & $\pm0.156\%$ \\
Quasi-QWERTY		& $14.453\%$ & $\pm0.155\%$ \\
Hooke				& $14.496\%$ & $\pm0.162\%$ \\
GK-S				& $14.866\%$ & $\pm0.148\%$ \\
GAG II				& $17.010\%$ & $\pm0.196\%$ \\
Lewis et al.			& $17.752\%$ & $\pm0.197\%$ \\
OPTI II				& $17.877\%$ & $\pm0.196\%$ \\
Dvorak				& $21.878\%$ & $\pm0.223\%$ \\
\hline
DGHP				& $6.920\%$ & $\pm0.123\%$ \\
\hline
\end{tabular}
\caption{The gesture recognition error rate and its associated statistics uncertainty for various
existing virtual keyboard layouts. The keyboards are ordered from best to worst in terms of their
error rates. At the very bottom is the error rate for the DGHP keyboard that was presented in Section \ref{sub:Keyboard-Optimization}.
It should be noted that the error rate for QWERTY and for DGHP are not identical to those quoted in the previous section because the keyboard geometries were modified to match the keyboard layouts in \cite{fastSwiping} as closely as possible.}
\label{table:keyboard_evaluations}
\end{table}

We can see that the results vary widely, which is expected given the drastic differences between some of the keyboard layouts.
The first thing to note is that the top two performing keyboards - DGHP and GK-C - were the only two keyboards that were optimized for gesture input clarity or gesture recognition error rate and show a reduction in error rate compared to QWERTY by $50.9\%$ and $35.6\%$, respectively.
The fact that DGHP outperforms GK-C is also expected since DGHP was optimized using the exact metric used in this evaluation.
The SWRM keyboard came in third even though this keyboard was optimized for input clarity with respect to tap input, not gesture input.
The GK-D and GK-T keyboards were optimized for gesture clarity but they were simultaneously optimized for one and two other performance metrics, respectively, so it is not surprising that they came in fourth and fifth.
Although the difference between the two is not statistically significant, the order they appear is exactly what we would expect. 

The worst keyboard by far is Dvorak which has an error rate that is $62.6\%$ higher than QWERTY.
We can also see that the hexagon keyboards tend to perform better than the square keyboards, although this comparison is not necessarily accurate given the subtleties mentioned in the previous paragraph.
It also interesting to see that just using an alphabetic keyboard will give you a much lower error rate than the majority of the keyboards listed in Table \ref{table:keyboard_evaluations}.

\section{\textbf{Discussion}}\label{sec:Discussion}
The results presented here show a clear demonstration of our proposed methodologies effectiveness with respect to calculating gesture recognition error rates and performing keyboard optimizations.
In contrast to recent work, such as the gesture clarity optimization performed by Smith, Bi, and Zhai, this new approach allows for the direct estimation of the error rate [\cite{googleKeyboard}].
This distinction may appear subtle, but it allows for ambiguities other than nearest neighbors to be taken into account.
These secondary ambiguities have a sizable impact on the real-world performance of any particular keyboard.
Additionally, threshold effects are more realistically taken into account with the reconstruction error rate estimation.
Words that have distant nearest neighbors will all have reconstruction rates that are effectively $100\%$.
Increasing the distance between these already distant words will increase the gesture clarity metric while having no effect on the actual reconstruction error rate.
In an optimization setting, these ineffectual changes might be preferred at the expense of making words with closer nearest neighbors more ambiguous, resulting in keyboards that are less effective overall.
Although these changes are significant, we consider the primary advancement of this methodology to be its ability to extend even further in quantifying metrics that accurately reflect the performance of different keyboards.
In this section, we will discuss some of the directions where this work and the Dodona framework can be built upon in future research.

One of the key advantages of the Monte Carlo evaluation approach is that any desired metric can be computed.
In the Smith, Bi, and Zhai paper, linear combinations of gesture clarity and gesture speed are optimized and then Pareto optimal configurations are found.
This is effective for finding keyboards that have relatively high gesture clarity and gesture speed metrics but requires making assumptions about the relative importance of each metric when choosing a keyboard that is optimal overall.
In past work, an optimal keyboard has typically been chosen from along the Pareto front under the assumption that each metric has equal importance [\cite{Dunlop:2012:MPO:2207676.2208659}, \cite{Xiaojun:Complete}, \cite{googleKeyboard}].
While this is a reasonable approach, there are certain cases where two metrics can be combined to more directly measure the overall performance.
This is the case with gesture clarity/error rate and gesture speed since the overall goal of minimizing the error rate is to speed up gesture input by minimizing the time required to correct mistakes.
With the Monte Carlo approach it is possible to do this by estimating the time it takes to input a gesture and then adding the correction time when a word is incorrectly reconstructed.
For example, the gesture input time could be estimated with the model used in \cite{googleKeyboard} and the correction time could be estimated by adapting the approach used in \cite{Arif} for gesture input.
This would allow for the direct estimation of how many words per minute can be entered on any given keyboard layout, leading to a metric that intuitively relates much more closely to how effective that layout would be in real-word usage.
The ability to reduce gesture speed and error rate into a single metric has the additional benefit of reducing the problem to one of scalar, rather than vector, optimization.
Roughly speaking, this allows you to focus the search in the direction that will directly maximize the underlying desired metric instead of optimizing in many different directions to extend the Pareto front (as was done in \cite{googleKeyboard}).
This significantly reduces the computational requirements of the optimization procedure.
Extensions like this are another possible area for future work and can be made very easily with the Dodona framework while adding virtually nothing to the overall computation time.

Another major strength of this approach is that it can be applied to an extremely wide variety of input methods with only the input models, and possibly the keyboards, needing to be modified.
Detailed discussion is beyond the scope of this paper, but the Dodona framework includes a model for touch typing which can be used to evaluate autocorrect on traditional keyboards or disambiguation in text entry on a T9 touchpad (on a touchpad keyboard which is also included in the framework).
New keyboard designs that use any sort of touch or gesture typing, with or without disambiguation, can be evaluated using the same overall methodology and framework.
This allows for extremely quick prototyping and testing of novel text input methods.

Reconstruction algorithms can also be easily modified to reflect more state of the art techniques.
Instead of considering individual word frequencies, word pair frequencies could be used to generate pairs of words and the previous word could modify the prior probabilities of the reconstructed words.
This would more closely mimic the behavior of commercially available gesture typing software and allow for more realistic estimation of metrics.
The algorithms could also be extended to include forgiveness of certain types of mid-gesture mistakes or allowing for common spelling mistakes.
This degree of flexibility is something that can only be achieved in a framework where actual reconstruction is taking place instead of relying on heuristic metrics.

One of the disadvantages of the Monte Carlo methodology is that it depends on the accuracy of the model.
A key improvement that could be made in future analyses is to utilize user data to tune and validate the input models.
The input models could also be extended to include more subtle aspects of user behavior such as mid-gesture mistakes and misspellings.
With well trained models, user behavior could be incorporated that would otherwise be impossible to take into account.
This would additionally make the exact values of the metrics meaningful while the untrained models used in this paper can only be used for relative keyboard comparisons.

One way this could be done is to incorporate touch accuracy into the gesture input models.
The model presented in this paper assumes that the accuracy of each key is identical.
However, Henze, Rukzio, and Boll used a crowd sourced dataset to show that tap accuracy is systematically skewed on mobile devices [\cite{TouchAccuracy}].
Furthermore, they showed that the frequency of touch errors for a specific target is correlated with the absolute position of the target on the touchscreen.
This could lead one to believe that certain keys are more susceptible to touch, and possibly gesture, inaccuracies than others.
Therefore, an obvious improvement to our user input model would be to systematically incorporate these results to adjust the accuracy at each location for a given keyboard geometry.
In addition, language model personalization could be incorporated and compared to the overall effect of keyboard layout on the error rate.
Fowler et al\mbox{.} showed that including language model personalization can dramatically reduce the word error rate for touch typing [\cite{Fowler}].

\section{\textbf{Conclusions}}\label{sec:Conclusions}

We have described a new way to model gesture input and
a procedure to evaluate the error rate of any touchscreen
keyboard layout. Using this method we have evaluated and compared the error rates for numerous
existing virtual keyboards and, specifically, shown that the QWERTY keyboard
is far from optimal for using a gesture input mechanism. We have also
described an optimization procedure for finding a keyboard that is better suited for gesture input.
We presented the most optimal keyboard that was found in our analysis
 and showed that it decreased the error rate for gesture input
by $50.1\%$ when compared to the QWERTY keyboard.

\section*{Acknowledgements}
We thank Duncan Temple Lang and Paul Baines of the Statistics
Department at UC Davis for allowing us access to the computing cluster
which provided the necessary computational power to run the final
optimizations. In addition, we thank Daniel Cebra and Manuel
Calderon de la Barca Sanchez of the Physics Department at UC Davis
for allowing us to make use of the Nuclear Physics Group computing 
resources while designing and testing the analysis software.

%\section*{References}
%\bibliographystyle{elsarticle-harv}
%\bibliography{dodona_ijhcs}

\end{document}